\documentclass[12pt]{article}

\usepackage{amssymb}
\usepackage{graphicx,psfrag}
\usepackage{subcaption}
\usepackage{enumerate}
\usepackage[superscript]{cite} 
\usepackage{url} 
\usepackage{hyperref}

\usepackage{longtable}

\usepackage[table]{xcolor} 
\usepackage{adjustbox}
\usepackage{booktabs}
\usepackage{pdflscape}
\usepackage{threeparttable}
\newcommand{\blind}{1}

\newcommand{\pkg}[1]{\texttt{#1}}
\newcommand{\fun}[1]{\texttt{#1}}

\def\iid{iid\ } 
\def\ie{i.e., }
\def\eg{e.g.\ }

\def\P{\text{P}} 
\def\E{\operatorname{E}} 
\def\I{\mathbf 1} 

\def\vecX{\mathbf{X}}
\def\vecx{\mathbf{x}}
\def\vecZ{\mathbf{Z}}
\def\vecz{\mathbf{z}}
\def\vecY{\mathbf{Y}}

\def\vecR{\mathbf{R}}
\def\D{\mathcal{D}}

\def\vecT{\mathbf{T}}
\def\vecM{\mathbf{M}}

\def\II{\mathbf{I}} 

\usepackage{amsthm}
\usepackage{amsmath}

\begin{document}

\def\spacingset#1{\renewcommand{\baselinestretch}%
{#1}\small\normalsize} \spacingset{1}


\if1\blind
{
  \title{\bf Bayesian additive regression trees for evaluating treatment
benefit predictors using observational data}
  \author{Yuan Xia, Mohsen Sadatsafavi, Jeenat Mehareen, and Paul Gustafson}\date{}
  \maketitle
} \fi

\if0\blind
{
  \bigskip
  \bigskip
  \bigskip
  \begin{center}
    {\LARGE\bf Evaluating Treatment Benefit Predictors using Observational Data: Contending with Identification and Confounding Bias}
\end{center}
  \medskip
} \fi

\bigskip

\begin{abstract}
A treatment benefit predictor (TBP) is an algorithm that maps a patient\textquotesingle s characteristics to their putative benefit from a given treatment, which can be used to inform treatment decisions.
However, a TBP must be evaluated in the target population before being adopted for patient care. 
When only observational data are available, evaluating TBPs requires standard causal identification assumptions, as treatment assignment in such settings is not random.
We obtain the posterior distributions of predictive performance measures to evaluate prespecified TBPs using observational data, by taking advantage of Bayesian additive regression trees (BART). 
We illustrate the evaluation of TBPs using selected measures and graphical visualizations: the concentration of benefit ($C_b$) index and the moderate calibration curve.
Simulation studies of binary and continuous outcomes settings, including balanced and imbalanced treatment allocation in the binary setting, establish the validity of the proposed approach.
In a case study, we use this approach to assess a TBP for systemic antibiotic therapy for patients with chronic obstructive pulmonary disease (COPD). 
We show that the constructed TBP does not in fact make calibrated predictions, because it both makes optimistic predictions of risks and exaggerates the risk reduction due to treatment. We conclude that flexible Bayesian approaches have the potential to assess TBPs, offering opportunities for flexible model specifications, adjustment for confounding, and uncertainty characterization.
\end{abstract}

\noindent%
\vfill

\newpage
\spacingset{1.45} 

\section{Introduction}
In clinical practice, it is becoming increasingly common to directly predict individualized treatment benefit to summarize the anticipated individual-level impact of an intervention within a target population.
We refer to any algorithm that maps a patient\textquotesingle s characteristics to their putative treatment benefit as a treatment benefit predictor (TBP).
Before TBPs are recommended for patient care, their predictive performance should be evaluated in the target population  \cite{riley2024evaluation}.
The predictive performance of TBPs can be measured in terms of overall, discrimination, calibration, and clinical utility, where discrimination and calibration can be viewed as components of overall measures (\eg mean squared error) \cite{spiegelhalter1986probabilistic}.
Discrimination assesses the monotonic association between the predicted and true quantities, whereas calibration assesses whether predictions agree with observations on average across the prediction range, rather than at the individual level.

For prespecified TBPs, Xia et al. (2025) exemplified the evaluation framework using observational data with a particular measure of discrimination, the concentration of benefit index ($C_b$), and a particular measure of calibration, the moderate calibration curve \cite{xia2025evaluating}.
The former partly reflects the ranking ability of the treatment benefit predictor, but it also captures the distribution of the predictions. 
The latter assesses whether the expected treatment benefit among individuals with the same predicted treatment benefit equals that predicted value.
We also regard these predictive performance measures as the target parameters and extend this work by studying how to estimate these parameters when observational data are used to represent the target population.

The target parameters for predictive performance evaluation can often be treated as a function of unknown distributional components, including nuisance functions such as the outcome process and the treatment assignment process \cite{Kennedy2016}.
Therefore, most existing methods for estimating these target parameters propose a plug-in estimator that is obtained by first estimating the nuisance functions and then plugging estimated nuisance functions into the identification formula defining the estimand.
Rather than using a single fitted nuisance-function value for each observation, as would result from classical maximum likelihood estimates of regression coefficients, and then obtaining a point estimate of the target quantity with uncertainty quantified using asymptotic normal inference or the nonparametric bootstrap \cite{abrevaya2015estimating, robertson2021assessing}, we consider a Bayesian approach.
In particular, we estimate performance measures of a prespecified TBP and quantify the uncertainty by obtaining their posterior distribution through Bayesian additive regression trees (BART).

BART, proposed by Chipman et al. (2010) \cite{chipman2010bart}, is a Bayesian nonparametric sum-of-trees model, which treats unknown target function(s) as parameters and takes draws from the posterior distribution of the parameters.
We consider BART not only for its ability to quantify uncertainty, but also for its flexibility in modeling the relationship between observed outcome, treatment, and covariates. 
In particular, it does not require researchers to choose which treatment-by-covariate interaction terms to include, since these are selected in a data-driven way.

We present this work in the following sequence. 
In Section 2, we define causal quantities for treatment benefit prediction and review the target estimands for illustrative predictive performance measures, and required assumptions.
Section 3 reviews the Bayesian additive regression tree (BART) method to prespecified TBP evaluation and describes how posterior samples of these predictive performance measures are obtained.
The proposed evaluation method is then illustrated in simulations in Section 4 and applied to a cohort of newly diagnosed patients with chronic obstructive pulmonary disease (COPD) from electronic health records in Section 5.

\section{Treatment Benefit Prediction and Evaluation}
Each individual in the target population is described by $(Y^{(0)}, Y^{(1)}, \vecX, \vecZ)$ with joint distribution $\mathbb{P}$.
For an active treatment of interest, $A$ denotes assignment of the treatment at a single time-point (not treated: $A = 0$ versus treated: $A = 1$), and $Y^{(a)}$ denote the outcomes that would be observed under $A = a$ for every patient in the population.
The vector $\vecX$ consists of pre-treatment covariates that will be used to predict treatment benefit in routine clinical practice, i.e., potential effect modifiers.
For evaluation, $\vecX$ must also be observable in the observational studies, which may provide a mix of predictive variables and confounders. 
The vector $\vecZ$ contains additional covariates that, together with $\vecX$, are required to control for confounding, but $\vecZ$ need not be effect modifiers and need not be available in routine practice.
Throughout this work, uppercase letters denote random variables, lowercase letters denote their observed realizations, and boldface letters denote objects more complex than scalars.

Let $B := (2s-1)(Y^{(1)} - Y^{(0)})$ be the random variable whose realizations are individual treatment benefits in the target population, where $s \in \{0,1\}$ indicate whether higher outcome values are favorable $(s=1)$ or adverse $(s=0)$.
As $B$ is unobservable, the ideal quantity to guide treatment decisions in clinical practice for an individual with $\vecX=\vecx$ is still the treatment benefit 
$$\tau_0(\vecx) = \E[B \mid \vecX = \vecx],$$
which conditions only on the routinely accessible covariates $\vecX$.
Denote the mean treatment benefit for smaller subgroups partitioned by both $\vecX$ and $\vecZ$ as $$\tau(\vecx, \vecz) = \E[B \mid \vecX = \vecx, \vecZ = \vecz].$$
Since $\vecX$ may represent only selected covariates that are feasible to ascertain during routine practice, treatment benefit often cannot be identified from $\vecX$ alone when identification requires conditioning on additional covariates. 
Quantities such as $\tau_0(\vecx)$ generally needs to be recovered from $\tau(\vecx,\vecz)$ by integrating over the conditional distribution of $\vecZ$ given $\vecX$.

Observed data from the observational study are realizations of a vector of random variables drawn from an underlying probability distribution $\mathbb{P}_{obs}$, denoted $(Y,  A, \vecX, \vecZ) \sim \mathbb{P}_{obs}$. 
We define a TBP as any function that takes $\vecx$ as input and  aims to predict $\tau_0(\vecx)$.
A prespecified TBP is denoted as $h(\vecx)$ and $H:= h(\vecX)$ as the random variable representing its predictions in the population.
Note that $\mathbb{P}_{obs}$ is a consequence of $\mathbb{P}$, as the observed outcome has the form $Y = Y^{(1)}A + Y^{(0)}(1-A).$
We define two conditional mean outcome functions as
\begin{align*}
    \mu(a, \vecx, \vecz) &= \E[Y \mid A =a, \vecX = \vecx, \vecZ = \vecz],\\
    \mu_0(a, \vecx) &= \E[Y \mid A =a, \vecX = \vecx].
\end{align*}
We denote the propensity score function as 
$$e(\vecx, \vecz) = \P(A = 1 \mid \vecX = \vecx, \vecZ = \vecz).$$

The following assumptions are commonly required, though they may not all be strictly necessary in all contexts \cite{greenland2017and}:
(1) no interference: between any two individuals, the treatment taken by one does not affect the potential outcomes of the other;
(2) consistency: the potential outcome under the observed treatment assignment equals the observed outcome $Y$; (3) overlap (positivity): the conditional probability of receiving the active treatment is bounded away from $0$ and $1$, i.e., $0 < e(\vecx, \vecz) < 1$, for all possible $\vecx$ and $\vecz$; and (4) conditional exchangeability: the treatment assignment is independent of the potential outcomes given the variables in $(\vecX, \vecZ)$. Therefore, we assume that the variables in $(\vecX, \vecZ)$ constitute a sufficient adjustment set to identify the treatment benefit.

\subsection{Predictive Performance Estimand}

Among possible measures, we illustrate how TBPs can be evaluated using $C_b$ index and the moderate calibration curve. 
For a given $h(\vecx)$, the estimand of $C_b$ is
\begin{align}
 C_b = 1 - \frac{\E[B]}{\E[B\eta(H)]} = 1 - \frac{\E[\tau(\vecX, \vecZ)]}{\E[\left(\tau(\vecX, \vecZ)\right)\eta(H)]},
\label{eq:cb_estimand_obs}
\end{align}
where $\eta(H) = 2F_H(H) - f_H(H)$.
Note that $F_H(\cdot)$ is the cumulative distribution function (CDF) of $H$, and $f_H(\cdot)$ is the probability mass function (PMF) of $H$.
When $H$ is continuous, $\eta(H) = 2F_H(H)$.
We have $0 \leq C_b < 1$ when $\E[B] > 0$ and $h(\vecx)$ is at least no worse than random.
As an extension of the Gini index, values of $C_b$ close to 1 indicate strong discriminatory ability of $h(\vecx)$, whereas values close to 0 indicate poor discrimination.

Similarly, we express the moderate calibration curve as
\begin{align}
 \E[B \mid H = h] = \E[\tau(\vecX, \vecZ)\mid H = h].
 \label{eq:cali_estimand_obs}
\end{align}
The $h(\vecx)$ is moderately calibrated when $\E[B \mid H = h]$ is on the 45-degree line in the calibration plot.
Note that $\tau(\vecx, \vecz)$ plays a vital role in determining both $C_b$ and $\E[B \mid H = h]$.
Under the listed assumptions,  $\tau(\vecx, \vecz)$ is identifiable using $\mathbb{P}_{obs}$ through
$$\tau(\vecx, \vecz) = (2s-1) \left(\mu(1,\vecx, \vecz) - \mu(0,\vecx, \vecz) \right),$$
where $s$ is known and fixed once the treatment benefit prediction problem is defined.

\section{Estimating Predictive Performance Measures}
In this section, we show how to obtain the posterior distributions of a scalar target parameter ($C_b$) and a function-valued parameter (the moderate calibration curve) of prespecified TBPs.
In particular, the posterior distributions of these parameters are derived from the posterior distribution of $\tau(\vecx,\vecz)$, where posterior samples of $\tau(\vecx,\vecz)$ are obtained using BART.

Under squared error loss, the optimal posterior summaries for $C_b$ and the moderate calibration curve are their respective posterior means, if they exist.
In particular, the posterior distribution of a parameter may be extremely skewed and heavy-tailed, and in some cases, the posterior mean may be infinite. 
In such cases, the posterior mean and related mean-based summaries no longer provide a reliable description of the parameter.
Other distributional summaries, such as median and quantiles, are more informative.

\subsection{Bayesian additive regression trees}
We briefly review BART for binary outcomes and explain what BART outputs, how those outputs are produced, and how they are used to draw posterior samples of $\tau(\vecx,\vecz)$.
A detailed version is provided in Appendix~A.
The posterior distribution is derived from the distribution of the sum of trees.
For binary outcome $Y_i$, Chipman et al. (2010) \cite{chipman2010bart} extended the development of BART for continuous outcomes using a probit latent variable model:
\begin{align*}
    Y^*_i &= g(a_i, \vecx_i,\vecz_i) +  \varepsilon_i, \quad \varepsilon_i \overset{\iid}{\sim} \text{N}(0, 1),\\
    Y_i &= \I(Y^*_i > 0),
\end{align*}
where $\I(\cdot)$ is the indicator function, $i = 1, 2, \cdots, n$, and $Y^*_i$ is an augmented latent variable with unit variance. Then, all $Y_i$ are independent Bernoulli random variables with conditional mean $\mu(a_i, \vecx_i,\vecz_i) = \Phi(g(a_i, \vecx_i,\vecz_i))$, where $\Phi(\cdot)$ is the standard normal CDF.  
BART models the function $g$ as a sum of trees,
\[g(a_i, \vecx_i,\vecz_i) = \sum^m_{j=1} g_j(a_i, \vecx_i,\vecz_i; \vecT_j, \vecM_j), 
\]
where $m$ is a fixed finite number of trees and each component $g_j$ is a tree-based function. The $j$-th tree-based function $g_j$ maps an input $(a_i, \vecx_i,\vecz_i)$ to a leaf value, with the destination determined entirely by its structure $\vecT_j$ (splitting rules) and terminal-node parameters $\vecM_j$ (leaf values). 

After a prespecified burn-in, we retain $l$ iterations, yielding an autocorrelated sample from the posterior of $g$. For binary outcomes with a probit link, we obtain posterior draws of the $\tau$ function, denoted as $\tau^{(1)}, \tau^{(2)}, \cdots, \tau^{(l)}$, by the deterministic transformation 
\begin{align*}
\tau(\vecx_i,\vecz_i) &= (2s-1)\left(\mu(1,\vecx_i,\vecz_i) - \mu(0,\vecx_i,\vecz_i)\right)\\
&=(2s-1)\left(\Phi(g(1, \vecx_i,\vecz_i)) - \Phi(g(0, \vecx_i,\vecz_i))\right).
\end{align*}
The Markov chain Monte Carlo (MCMC) draws from the posterior distribution of $\tau$ are as follows:
\begin{align*}
     \tau(\vecx,\vecz) = \begin{pmatrix}
\tau^{(1)}(\vecx_1, \vecz_1) & \tau^{(1)}(\vecx_2, \vecz_2)& \cdots & \tau^{(1)}(\vecx_n,\vecz_n)\\
\tau^{(2)}(\vecx_1,\vecz_1) & \tau^{(2)}(\vecx_2,\vecz_2)& \cdots & \tau^{(2)}(\vecx_n,\vecz_n)\\
\vdots & \vdots & \vdots & \vdots\\
\tau^{(l)}(\vecx_1,\vecz_1) & \tau^{(l)}(\vecx_2,\vecz_2)& \cdots & \tau^{(l)}(\vecx_n,\vecz_n)
\end{pmatrix},
\end{align*}
where the $i$-th column contains the $l$ posterior draws of $\tau$ evaluated at the observation $(\vecx_i,\vecz_i)$, and the $l$-th row contains the $\tau^{(l)}$ evaluated at all observations in $\D$.

\subsection{Concentration of Benefit Index}
According to the estimand of $C_b$ in Equation~(\ref{eq:cb_estimand_obs}), the uncertainty in the estimate of $C_b$ comes from two sources: the function $\tau$ and the joint distribution $F_{\vecX,\vecZ}$. 
The posterior distribution of $\tau$ is induced by the posterior distribution of $g$, since $\tau$ is a function of $g$.
We use Bayesian bootstrap to inject the uncertainty about the population distribution of $(\vecX,\vecZ)$ through Dirichlet weights $\left(\omega_1, \cdots, \omega_n \right) \sim \text{Dir}(1,\cdots, 1)$. 
Since the class of distribution functions is restricted to 
\begin{align*}
    F_{\vecX,\vecZ}(t,s) = \sum^n_{i=1} \omega_i \I(\vecx_i \leq t, \vecz_i \leq s),
\end{align*}
the posterior distribution of $F_{\vecX,\vecZ}$ is obtained as the pushforward of the Dirichlet posterior under this mapping.
In the absence of prior information suggesting dependence between $g$ and $F_{\vecX,\vecZ}$, it is reasonable to assign independent priors to them.
Combined with the factorization of the likelihood, this implies posterior independence between $\tau$ and $F_{\vecX,\vecZ}$.
Given the joint posterior distribution of $\tau$ and $F_{\vecX,\vecZ}$, the exact posterior of $C_b$ is the pushforward of the joint posterior distribution of $\tau$ and $F_{\vecX,\vecZ}$.
Thus, given $h(\cdot)$ and posterior draws of $\tau(\vecx, \vecz)$ and $F_{\vecX,\vecZ}(\vecx, \vecz)$, we can obtain posterior draws of $C_b$.

The ratio of two means in the formula for $C_b$ is estimated using the ratio of Bayesian bootstrap means. The $l$-th posterior draw of $C_b$ can be expressed as 
\begin{align}
    C_{b}^{(l)} =1 - \frac{\sum^n_{i=1}  \omega^{(l)}_i \cdot \tau^{(l)}(\vecx_i,\vecz_i)}{\sum^n_{i=1}  \omega^{(l)}_i \cdot  \tau^{(l)}(\vecx_i,\vecz_i) \cdot \eta^{(l)}(h(\vecx_i))},
    \label{eq:cb_est}
\end{align}
where $\omega^{(l)}_i$ denotes the $l$-th posterior draw of the Dirichlet weight for unit $i$ and $\eta^{(l)}$ is the $l$-th posterior draw for $\eta$.
If $H$ is continuous, we have 
\begin{align}
\eta^{(l)}(h(\vecx_i)) =  2\sum^n_{i'=1} \omega^{(l)}_{i'} \I(h(\vecx_{i'}) \leq h(\vecx_i)).
\label{eq:eta_est_continous}
\end{align}
If $H$ is discrete, we need an additional correction term; thus, we have 
\begin{align}
\eta^{(l)}(h(\vecx_i)) =  2\sum^n_{i'=1} \omega^{(l)}_{i'} \left( \I(h(\vecx_{i'}) \leq h(\vecx_i)) - \frac{1}{2}\I(h(\vecx_{i'}) = h(\vecx_i)) \right).
\label{eq:eta_est_discrete}
\end{align}

The following caveat for the estimation of $C_b$ should be taken into account.
Recall that $C_b$ of $h(\vecx)$ can be viewed as a functional defined in terms of a pair of functions $\tau$ and $F_{\vecX,\vecZ}$.
The $C_b$ lies in $[0,1)$ under either of two conditions: $0 < \E[B] \leq \E[B\eta(H)]$ or $\E[B\eta(H)] \leq \E[B] < 0$.
Since our focus is primarily on settings in which the average treatment benefit in the population is positive, we require $h(\vecx)$ to be at least not worse than random.
However, posterior draws of $\tau$ do not guarantee that every corresponding value of $\E[B]$ computed from each draw is positive; meanwhile, the posterior draws of $F_{\vecX,\vecZ}$ do not guarantee that $\E[B] \leq \E[B\eta(H)]$.
Therefore, it is possible for the posterior draws of $C_b$ to be outside the range $[0,1)$, especially when the treatment benefits are small and the sample data provide limited information about the treatment benefits.
Yet, we are not aware of a straightforward remedy for this issue without introducing additional assumptions.

\subsection{Moderate Calibration Curve}
For a moderate calibration curve, the estimand is defined in Equation~(\ref{eq:cali_estimand_obs}).
If $H$ is continuous, estimating $\E[B \mid H = h]$ requires regularization to avoid an overly noisy curve.
We use the Nadaraya-Watson kernel smoothing estimator \cite{bierens1988nadaraya}, which can be expressed as
\begin{align*}
   \hat{\E}[ \tau(\vecX, \vecZ) \mid H = h] =  \frac{\sum^n_{i=1} K\left((h - h_i)/d_n\right) \tau(\vecx_i, \vecz_i)}{\sum^n_{i=1} K\left((h - h_i)/d_n \right)} ,
\end{align*}
where $K(\cdot)$ is a kernel smoothing function and $d_n$ is the bandwidth.
We use a kernel smoother based on a Gaussian kernel, so that observations with $H$ values closer to the target receive higher weights. The bandwidth parameter determines the effective neighborhood size, with smaller bandwidths producing tighter and more localized smoothing.
Since the calibration assessment depends on how closely the estimated curve follows the reference line, the bandwidth is set to a fixed fraction of the observed range of $H$, without cross-validation. The choice of bandwidth, although it affects visual smoothness, does not substantially affect the qualitative conclusion.
For the $l$-th posterior draw $\tau^{(l)}$, a smoothed curve is estimated, which can be expressed as 
\begin{align}
    \E^{(l)}[ B \mid H = h] = \frac{\sum^n_{i=1} K\left((h - h_i)/d_n\right) \tau^{(l)}(\vecx_i, \vecz_i)}{\sum^n_{i=1} K\left((h - h_i)/d_n \right)} .
    \label{eq:cali_est_continous}
\end{align}

If $H$ is discrete, we estimate the $l$-th posterior draw of the moderate calibration curve by the sample average of $\tau^{(l)}(\vecx_i, \vecz_i)$ within the subset of observations that have the same $H$ value, which can be expressed as 
\begin{align}
    \E^{(l)}[ B \mid H = h] = \frac{ \sum_{i=1}^n   \I(H_i=h) \tau^{(l)}(\vecx_i, \vecz_i)}{\sum_{i=1}^n \I(H_i=h)} .
    \label{eq:cali_est_discrete}
\end{align}
This estimator relies on having a sufficient number of observations in each stratum defined by the value of $H$; sparse strata may lead to unstable or unreliable estimates.

\section{Simulation Studies}
We perform simulation studies to demonstrate that BART (i), can indeed recover the target $\tau(\vecx,\vecz)$, and (ii), obtain the posterior distributions of the $C_b$s and the moderate calibration curves for prespecified TBPs.
The simulation studies draw observations from two of the three synthetic populations introduced in Xia et al. (2025), where the true evaluation results were established for prespecified TBPs. 
Because these evaluation results are known, we can directly assess how closely the estimated $C_b$ and the moderate calibration curve from the proposed method align with the population-level truths.
Note that these synthetic populations have $s = 1$ and $B := Y^{(1)} - Y^{(0)}$, where $B > 0$ denotes an improvement in the favorable outcome.

\textbf{Binary covariates and outcome.} We generate $n =$ 5,000 observations from $(Y, A, \vecX, Z) \sim \mathbb{P}_{obs,1}$, where $\mathbb{P}_{obs,1}$ arises from $\mathbb{P}_{1}$ and the treatment assignment mechanism:
\begin{align*}
    (Y^{(0)} \mid X_1 = x_1, X_2 = x_2, Z = z) &\sim \text{Bernoulli}(\alpha_{00} + \alpha_{01}x_1 + \alpha_{02}x_2 +\alpha_{03}z),\\
    (Y^{(1)} \mid X_1 = x_1, X_2 = x_2, Z = z) &\sim \text{Bernoulli}(\alpha_{10} + \alpha_{11}x_1 + \alpha_{12}x_2 +  \alpha_{13}z),\\
    (A \mid X_1 = x_1, X_2 = x_2, Z = z)  &\sim \text{Bernoulli}(\beta_0 + \beta_1z),\\
    (X_1, X_2, Z) &\sim \text{Multivariate Bernoulli}\big(p\big),
\end{align*}
where $p = (p_{111}, p_{110}, p_{101}, p_{100}, p_{011}, p_{010}, p_{001}, p_{000})$.
Let $\vecX = (X_1, X_2)$.
The exposure $A$ is a binary indicator of bronchodilator therapy, and the outcome $Y$ is a binary indicator of an improvement in forced expiratory volume in one second ($\text{FEV}_1$). 
The individual treatment benefit $B$ of having bronchodilator therapy can take values of $-1$, $0$, or $1$, where $-1$ indicates harm, $0$ indicates no benefit, and $1$ indicates benefit. 
The identifiable subgroup treatment benefit is
\begin{align*}
    \tau(x_{1},x_{2}, z) = \alpha_0 + \alpha_1x_{1} + \alpha_2 x_{2} + \alpha_3 z,
\end{align*}
where $(\alpha_0, \alpha_1, \alpha_2, \alpha_3) = (-0.2942, 0.1604, 0.1456, 0.3717)$.
We consider four prespecified TBPs: $h_1(\vecx)$, $h_2(\vecx)$, $h_3(\vecx)$, and $h_4(\vecx)$, which represent the covariate mean predictor, the moderately calibrated predictor, the strongly calibrated predictor, and the predictor subject to confounding bias due to incomplete control of confounding, respectively.
We consider two parameter settings to reflect real-world conditions: (1) balanced treatment allocation $(\beta_0,\beta_1)=(0.1204,0.7621)$, \ie the marginal probabilities of assignment to treatment and control are approximately $0.5$; and (2) unbalanced treatment allocation $(\beta_0,\beta_1)=(0.1204,-0.1)$, in which the number of treated units is substantially different from the number of untreated units.
Particularly, the marginal probability of receiving treatment is approximately $0.07$ in the unbalanced setting.
This reduction in probability is driven by a lower treatment rate among patients with $Z=1$.

\begin{figure}
    \centering
    \includegraphics[scale=0.75]{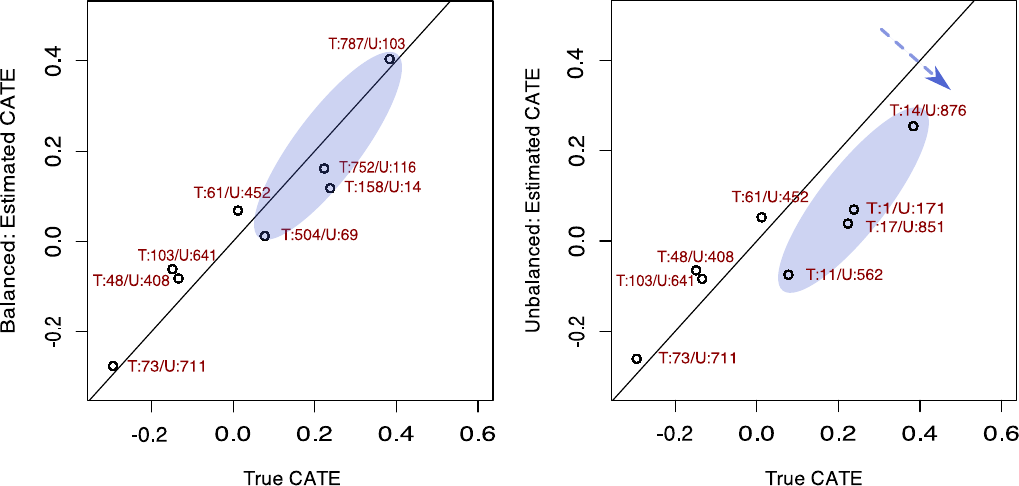}
    \caption{Scatterplot of the actual $\tau(x_1,x_2,z)$ plotted against the estimated $\tau(x_1,x_2,z)$ from BART for the first synthetic population. The text indicates the count of observed treated (T) and untreated units (U). The ellipse encircles the units with $Z = 1$, representing the group of patients with fewer treated units and more untreated units.}
    \label{fig:1}
\end{figure}

Figure \ref{fig:1} displays the relationship between the true population-level values of $\tau(x_1,x_2,z)$ (x-axis) and their corresponding posterior mean estimates (y-axis) in both treatment-allocation settings.
Each circle represents a posterior draw for each observed $(x_1, x_2, z)$ obtained using the default BART settings with $m = 200$ trees and 1,000 burn-in iterations.
With $l =$ 1,000 posterior draws evaluated at every observed point, these circles overlap heavily at each location in Figure \ref{fig:1}.
In the balanced setting, posterior mean estimates produced by default BART are close to the true $\tau(x_1,x_2,z)$ values. 
In contrast, in the unbalanced setting, the estimates for units with $Z=1$ tend to be smaller than the true $\tau(x_1,x_2,1)$.
This underestimation is not bias caused by BART, as the same model performs reasonably well in the balanced setting for this target population.
Rather, it is because the data contain little information about treatment benefit under the unbalanced treatment allocation, so any model is expected to exhibit reduced predictive accuracy.

\begin{figure}[h]
    \centering
    \includegraphics[scale=0.45]{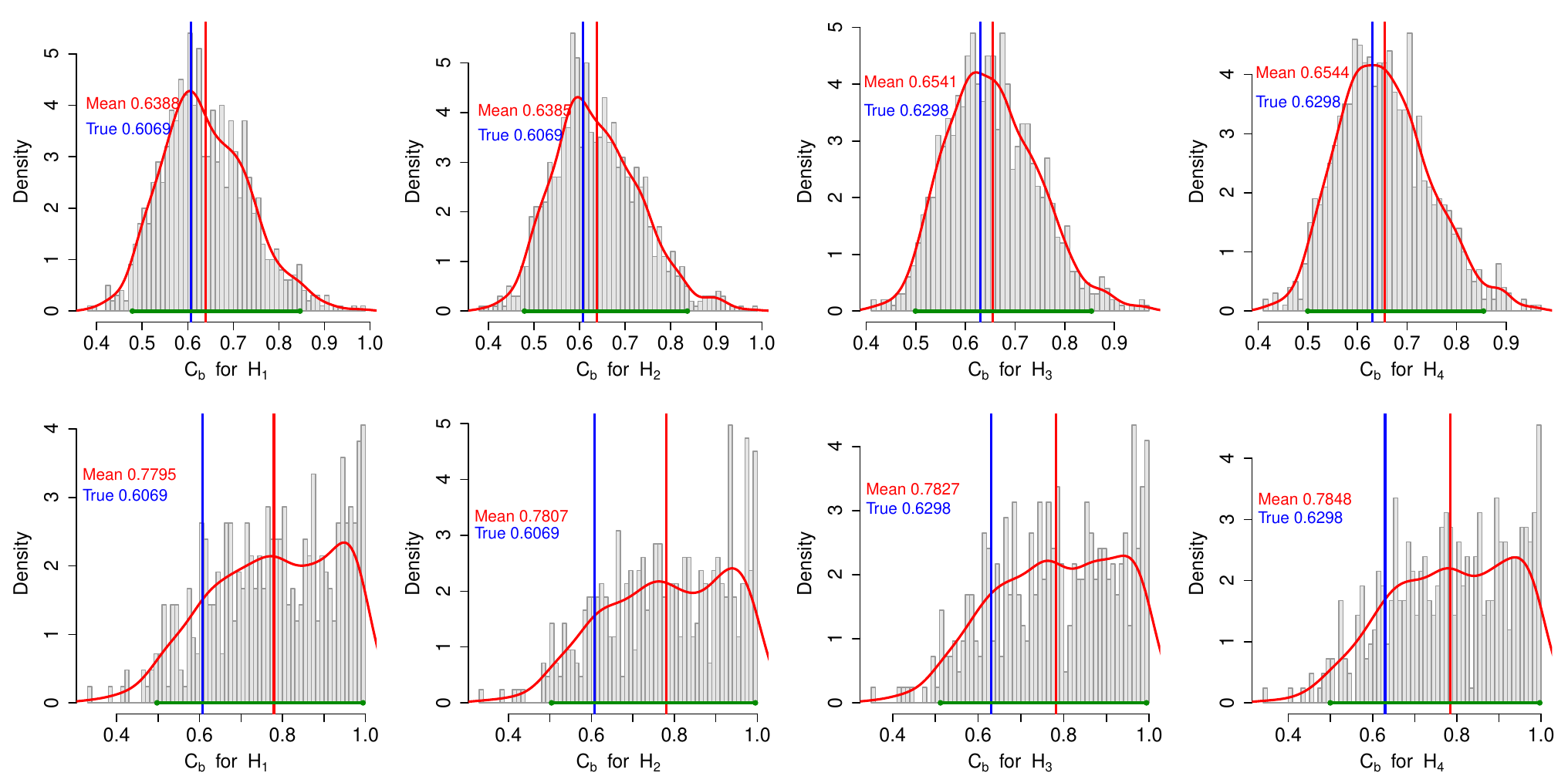}
    \caption{Histogram of estimated $C_b$ for TBPs, with red for mean, blue for actual, and green for 95\% credible interval. Top row: balanced treatment assignment; bottom row: unbalanced treatment assignment.}
    \label{fig:2}
\end{figure}

We estimate $C_b$ for the four TBPs using the estimator defined in Equation~(\ref{eq:cb_est}) and Equation~(\ref{eq:eta_est_discrete}).
This estimator performs well when treatment assignment is balanced.
Figure~\ref{fig:2} presents the posterior distribution of $C_b$, where the posterior mean of $C_b$ remains close to, but slightly overestimates, the true $C_b$ under balanced assignment. 
With an unbalanced assignment, the estimator can be less accurate, showing large variances and even values outside the defined range $[0,1)$. 
For clarity, Figure~\ref{fig:2} presents only those $C_b$ values within the range, and roughly half of the posterior draws in the unbalanced cases lie outside it.
Therefore, the $C_b$ estimator, as a ratio of sample averages, can be very sensitive under extreme treatment imbalance, where small denominator errors might be magnified.

\begin{figure}[h]
    \centering
    \includegraphics[scale=0.45]{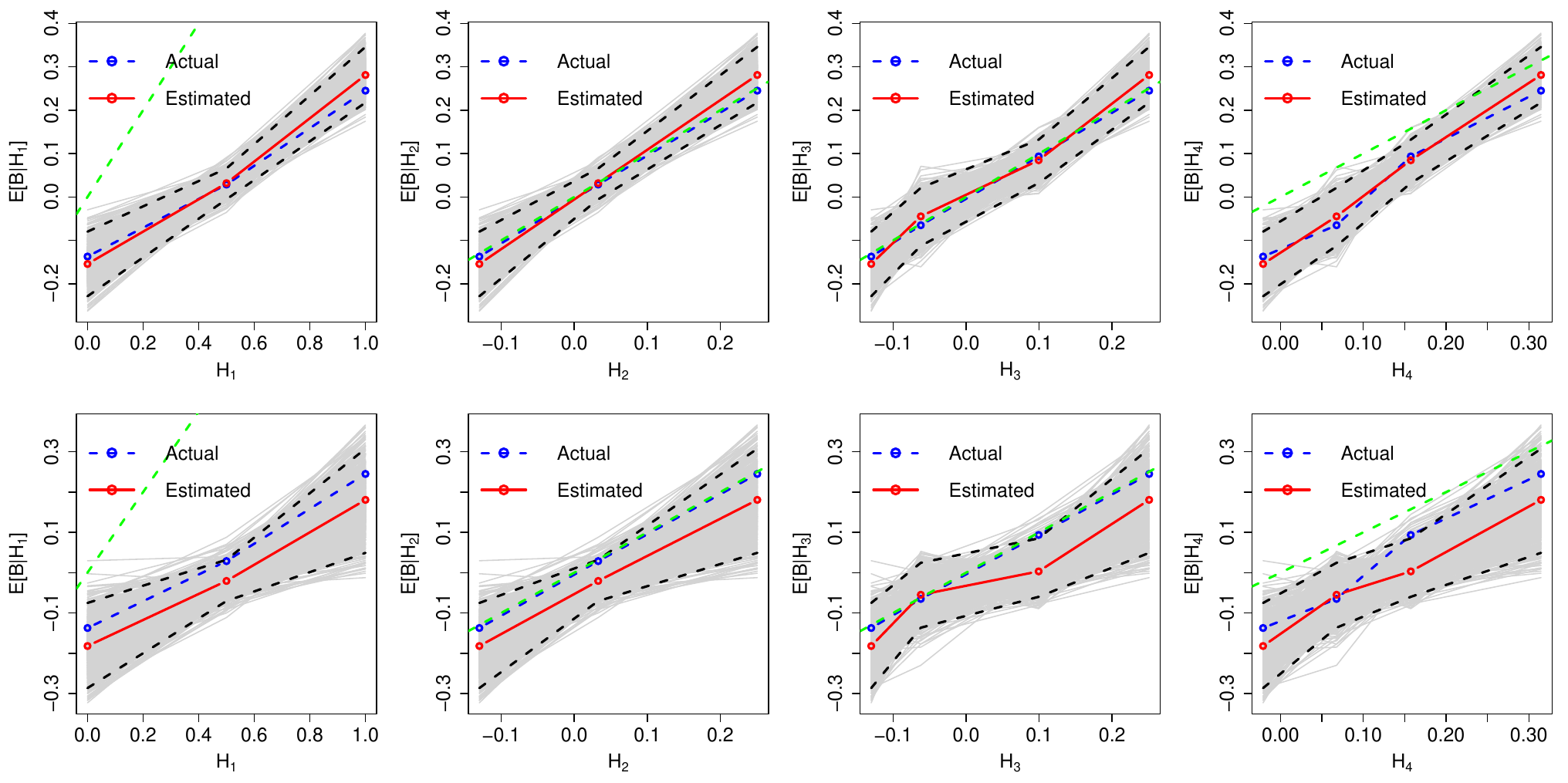}
    \caption{Estimated calibration curves for TBPs are in red, actual curves in blue. Black dashed curves mark the 95\% Bayesian credible interval boundaries, and the green line is the 45-degree reference line. Top row: balanced treatment assignment; bottom row: unbalanced treatment assignment.}
    \label{fig:3}
\end{figure}

We estimate the moderate calibration curves for the four TBPs using the estimator defined in Equation~(\ref{eq:cali_est_discrete}).
Figure~\ref{fig:3} shows the posterior draws (in gray), the posterior means (in red), and the actual moderate calibration curves (in blue) for the TBPs, in both balanced and unbalanced treatment-allocation settings.
Under balanced assignment, the red curves overlap the blue population-level truths. Under an unbalanced assignment, the posterior means are underestimates, but the truth still falls within the wider 95\% Bayesian credible intervals. 
This moderate calibration curve estimator has lower variance in a balanced case than in an unbalanced one. 
Overall, the moderate calibration assessment is affected by the unbalanced treatment allocation; nevertheless, it still conveys the evaluation in the correct direction.

\textbf{Continuous covariates and outcome.} Let $\vecX = (X_1, X_2)$. We evaluate the performance of $h(x_1, x_2) = x_1 + x_2$ using $n =$ 5,000 randomly sampled observations from $(Y, A, \vecX, Z) \sim \mathbb{P}_{obs,3}$, where $\mathbb{P}_{obs,3}$ is induced by $\mathbb{P}_{3}$ proposed by Foster and Syrgkanis (2023) \cite{foster2023orthogonal} and the treatment assignment mechanism:
\begin{align*}
    (Y^{(a)} \mid X_1 =x_1, X_2 = x_2, Z = z)  &\sim \text{N}(\tau(x_1, x_2, z)\left(a - 0.5\right) + b(x_1, x_2, z), \sigma^2),\\
    (A \mid X_1 = x_1, X_2 = x_2, Z = z) &\sim \text{Bernoulli}(e(x_1, x_2,z)),\\
    (X_1, X_2, Z) &\overset{iid}{\sim} \text{Unif}(0,1),
\end{align*} 
where $\sigma = 0.1$, the propensity score function $e(x_1, x_2, z) =z$, the base response function $b(x_1, x_2, z) = \max \left(x_2, z\right) + 0.1x_1$, and 
\begin{align*}
  \tau(x_1, x_2, z) &=\tau_0(x_1, x_2) =  \max \left(x_1, x_2\right).
\end{align*}
The treatment allocation is marginally balanced with $\P(A = 1) = 0.5$, though $e(\vecx,z) = z$.

\begin{figure}[h]
    \centering
    \includegraphics[scale=0.35]{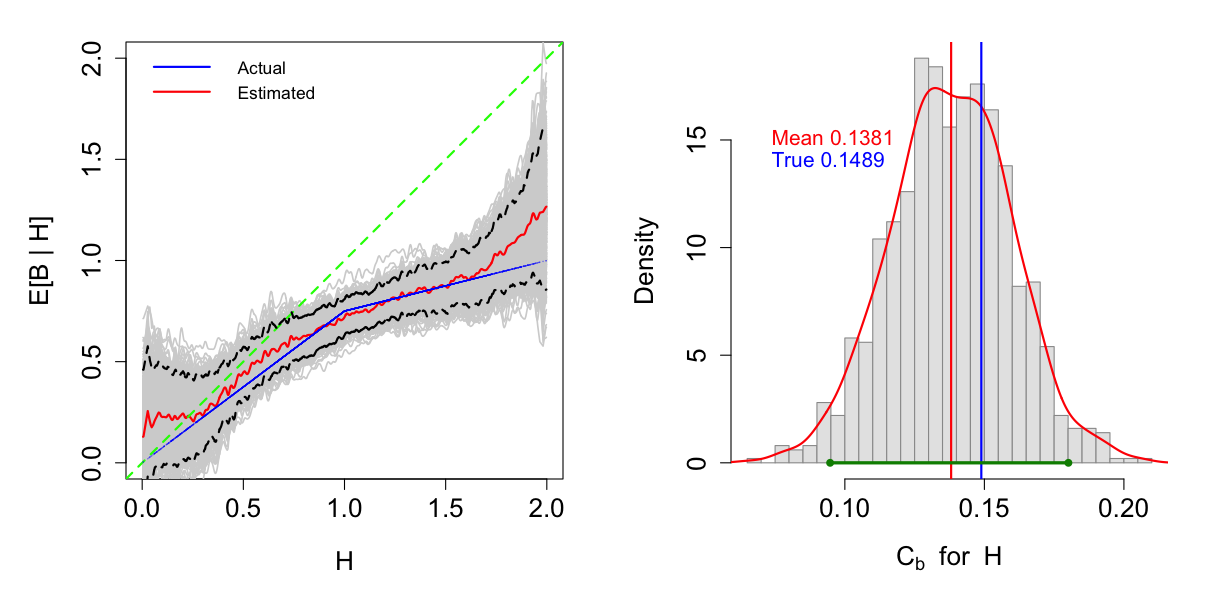}
    \caption{Moderate calibration plot (right) and posterior distribution of $C_b$ (left) for $h(x_1, x_2) = x_1 + x_2$. Left: the red curve represents the posterior mean of the moderate calibration curve, and the green dashed line indicates the 45-degree reference line. Right: the posterior mean of $C_b$ is shown in red, the true $C_b$ in blue, and the 95\% credible interval in green.}
    \label{fig:4}
\end{figure}

We estimate $C_b$ using the estimator defined in Equation~(\ref{eq:cb_est}) and Equation~(\ref{eq:eta_est_continous}), and estimate the moderate calibration curve using the estimator defined in Equation~(\ref{eq:cali_est_continous}).
Each smoothed moderate calibration curve is obtained from a kernel regression smoother with a Gaussian kernel and a bandwidth of $d_n = 0.2$. 
Figure~\ref{fig:4} presents the posterior distributions of the moderate calibration curve and $C_b$ for $h(x_1, x_2)$. 
The posterior mean of moderate calibration curve, implemented by averaging the 1,000 posterior draws of the moderate calibration curve, closely follows the actual moderate calibration curve, except near the boundaries, where the curves become wiggly and show large variability. 
For $C_b$, the posterior mean slightly underestimates the true $C_b$, but the 95\% credible interval captures it. 

Overall, the estimated measures closely agree with the true, closed-form results in both simulation studies. 
Although estimators for both $C_b$ and the moderate calibration curve are affected by treatment allocation, the estimator for the moderate calibration curve is substantially less sensitive to this imbalance than the estimator for $C_b$. 
Thus, in the next section, we estimate only the moderate calibration curve for a prespecified TBP using real data that exhibit an imbalanced treatment allocation.

\section{Applied Example: Evaluation of a TBP for Antibiotic Therapy to Prevent Exacerbations in COPD} \label{sec:copd}
COPD is a progressive respiratory condition characterized by airflow limitation and chronic inflammation of the airways \cite{calverley2023contemporary}. 
Patients with COPD often experience acute exacerbations, which are sudden episodes of worsening respiratory symptoms that accelerate lung function decline, impair quality of life, and increase mortality risk.
Typical preventive therapies for exacerbations are inhaled bronchodilators and corticosteroids. 
However, some patients do not respond adequately to such treatments. 
In these patients, treatment with azithromycin, a systemic antibiotic, has been shown to reduce the frequency of COPD exacerbations, and is therefore used as a preventive therapy in selected patients \cite{albert2011azithromycin}. 
However, the benefit of Azithromycin is not uniform across individuals, motivating efforts to predict who is most likely to benefit from the treatment.
In this study,  we first construct a TBP to predict the benefit of azithromycin on acute exacerbations.
We then aim to externally validate this TBP using real-world data from the UK\textquotesingle s clinical practice research datalink (CPRD) \cite{wolf2019data}, thereby assessing its calibration in an independent population.
We consider only the moderate calibration curve in this example, not $C_b$.
The specific database we used is CPRD Aurum, an anonymized primary care database of electronic health records spanning January 1, 2004, to January 31, 2024.
Ethics approval for this study was obtained from the UBC Research Ethics Board (H23-00752).

\subsection{Methods}
We evaluate a TBP on a study cohort from the target population. 
This study cohort consists of newly diagnosed COPD patients identified in CPRD Aurum, linked to mortality records and hospital admissions from January 1, 2004, to March 31, 2021.
It covers the period with overlapping coverage across the linked data sources.
In line with the inclusion criteria of Albert et al. (2011) \cite{albert2011azithromycin} and Adibi et al. (2020) \cite{adibi2020acute}, the study cohort is restricted to age $\geq$40 years at initial diagnosis; spirometry-confirmed COPD; and current or former smoking. 
We focus on COPD patients who experienced at least one exacerbation episode, and define the index date as the end of the first acute exacerbation occurring after the initial COPD diagnosis date. 

The treatment of interest is oral azithromycin. We identified oral azithromycin prescriptions in CPRD Aurum using product codes for tablet and capsule formulations. 
The outcome of interest is a composite of COPD exacerbation (moderate or severe) and all-cause mortality. 
A moderate exacerbation was defined in primary-care records as any of the following: (i), prescriptions for both an antibiotic and an oral corticosteroid for 5-14 days; (ii), two or more respiratory symptoms accompanied by a prescription for an antibiotic or an oral corticosteroid; (iii), a lower respiratory tract infection code; or (iv), an acute COPD exacerbation code. We defined a severe exacerbation as a hospitalization with ICD-10 codes J41\text{--}J44, indicating COPD exacerbation. Patients could experience multiple moderate or severe exacerbations. 
Exacerbations recorded within 14 days of each other were considered a single exacerbation episode.

\begin{figure}
    \centering
    \includegraphics[scale = 0.5]{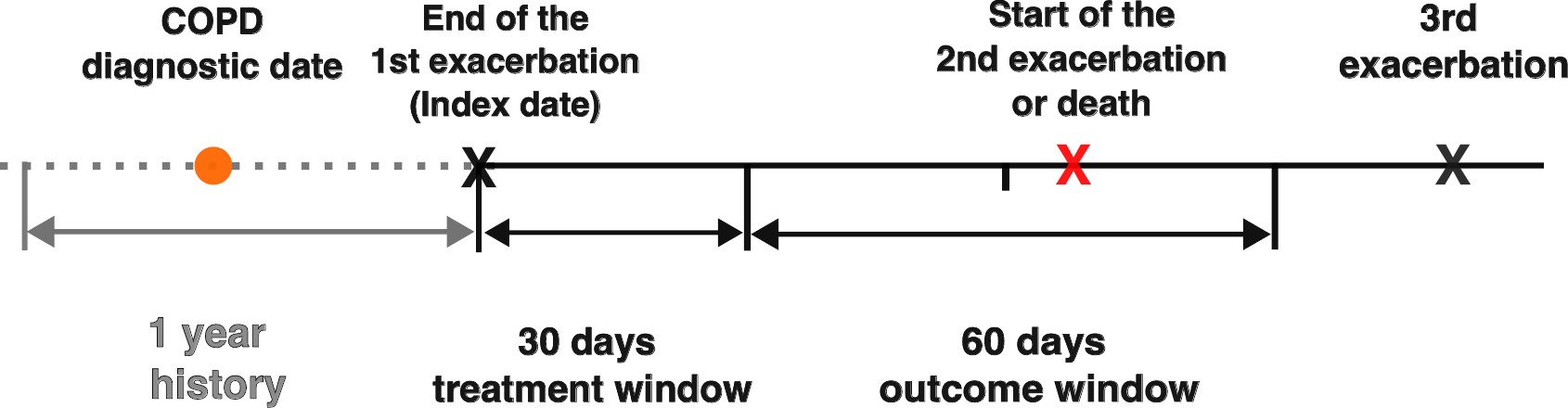}
    \caption{Cohort definition flow chart. The index date is shown as the black cross, which appears after the orange dot representing the initial COPD diagnosis date.}
    \label{fig:5}
\end{figure}

The cohort definition flowchart is shown in Figure~\ref{fig:5}.
We have $s = 0$ since the outcome is unfavorable.
We estimate the benefit of initiating oral azithromycin within 30 days of the index date versus no initiation during that window, evaluating the outcome over a fixed 60-day follow-up period. 
This corresponds to an intention-to-treat analysis in an observational setting \cite{olijnyk2022understanding}. 
Accordingly, real-life behaviors that occur after the treatment window, including subsequent adherence and treatment changes, are regarded as part of the overall effect.
Because the retrospective data provide almost complete 60-day follow-up after the index date, we treat the outcome as a binary 60-day outcome rather than time-to-outcome, and we do not explicitly model censoring in this intention-to-treat analysis.
If any azithromycin was prescribed during the 30-day treatment window, we set $A=1$; otherwise, $A=0$. 
If a second exacerbation or death occurred in the 60-day outcome window, $Y=1$; otherwise $Y=0$.
Moreover, we excluded patients who experienced an outcome within 30 days after the index date, as events occurring within this window were assumed to represent a resurgence of the prior exacerbation rather than a new episode.
To emulate the new initiation of azithromycin, we require a 12-month washout period with no azithromycin prescribed before the index date.

\textbf{The TBP to be evaluated:} 
The TBP was constructed to reflect the current published evidence on the use of azithromycin to prevent acute exacerbations.
Acute COPD exacerbation prediction tool (ACCEPT) proposed by Adibi et al. (2020) is an algorithm that predicts the individualized 12-month risk of acute COPD exacerbations from baseline characteristics $\vecX$ \cite{adibi2020acute}.
The vector $\vecX$ used in ACCEPT includes a history of exacerbations, age, sex, $\text{FEV}_1$, body mass index (BMI), smoking status, domiciliary oxygen therapy, lung function, symptom burden, and prior-year and current medication use.
The authors also suggested using this risk prediction model as a TBP for exacerbation risk reduction as a result of azithromycin therapy, by applying a fixed hazard ratio of  $0.73$ estimated from the MACRO study \cite{albert2011azithromycin}.
The value of $0.73$ indicates a $27\%$ lower hazard with the daily oral azithromycin compared to usual care.
Since ACCEPT predicts the risk of exacerbation at 12 months while the outcome window of interest is only 2 months, the constructed TBP of azithromycin on the outcome $Y$ that we consider is
\begin{align*}
h(\vecx) = \left(\left(1 - \pi_{12}(\vecx)\right)^{1/6}\right)^{0.73} - \left(1 - \pi_{12}(\vecx)\right)^{1/6},
\end{align*}
where $\pi_{12}(\vecx)$ is the predicted probability of having at least one exacerbation within 12 months, without azithromycin, from ACCEPT.
The probability of having no outcome $Y$ within 2 months without azithromycin is estimated by $\left(1 - \pi_{12}(\vecx)\right)^{1/6}$, assuming a constant hazard over time.
Appendix~B provides a more detailed summary of these two related studies and describes in detail how their findings were integrated to obtain this TBP.

Because of potential confounding by indication and disease severity, adjustment for confounders is necessary.
In addition to the vector $\vecX$, additional confounding variables, denoted by $\vecZ$, were identified through a review of the relevant literature and consultation with clinical experts. 
These included 14 variables related to concomitant medications, comorbidities, and other patient characteristics.
Concomitant medications used after index date and before azithromycin initiation are summarized as a binary indicator of any concomitant medication use.
Comorbidities were identified from the 12 months preceding the index date.
Vector $\vecZ$ also includes socioeconomic status, the latency between the COPD diagnosis date and the index date, and the scaled calendar year of the index date. 
Socioeconomic status proxies access and affordability of oral azithromycin; latency serves as a marker of COPD severity; and scaled calendar time adjusts for contemporaneous prescribing policies and practice patterns.
See Appendix~C for detailed reasoning and evidence on the construction of $\vecZ$.

We assume that the conditional exchangeability assumption holds after adjusting for both $\vecX$ and $\vecZ$.
We also assume that no interference and consistency assumptions hold, ensuring identification of the treatment benefit.
The assumption of no interference appears reasonable in this context, as initiation of azithromycin in one patient is unlikely to affect the occurrence of exacerbation in another.
The assumption of consistency is also plausible because treatment is well defined as the initiation of oral azithromycin within a fixed window, among eligible COPD patients, with exacerbation measured over a prespecified follow-up period.

Having specified $\vecX$ and $\vecZ$, we next consider missing data.
Among all identified patient characteristics in $\vecX$ and $\vecZ$, four contained missing values. 
We assume these variables are missing at random and therefore handle them using separate imputation methods, since the current BART implementation does not seamlessly integrate imputation within the MCMC.
Specifically, predictive mean matching was used for the numeric variables, and proportional-odds logistic regression was used for ordered categorical variables. 
We use an iterative imputation procedure in which missing values for each variable are filled in using draws from the observed data.
Given the illustrative nature of the example, we perform a single imputation rather than multiple imputations for computational simplicity.
See Appendix~D for details on missing data and the imputation procedures used.

\subsection{Results}

\begin{table}
\centering
\caption{Baseline characteristics of study cohort.\label{tab:1}}%
\small
\setlength{\tabcolsep}{6pt}
\renewcommand{\arraystretch}{0.5}
\begin{adjustbox}{width=\linewidth}
\begin{tabular}{llll}
\toprule
 Characteristic & No azithromycin & Azithromycin  & Vector\\
 \midrule
Age (mean (SD)) & 66.07 (11.39) & 66.35 (11.62)& $\vecX$\\
Female sex (\%) & 121,484 (46.8) & 216 ( 49.1) & $\vecX$\\
Current smoker (\%) & 144,892 (55.8) & 216 ( 49.1) & $\vecX$\\
BMI (mean (SD)) & 27.25 (6.25) & 27.64 (6.74)& $\vecX$\\
IMD$\_5$ (\%) &  & & $\vecZ$\\
\quad Quintile 1 & 75,055 (28.9) & 95 (21.6)&\\
\quad Quintile 2 & 57,297 (22.1) & 97 (22.0)&\\
\quad Quintile 3 & 47,796 (18.4) & 86 (19.5)&\\
\quad Quintile 4 & 44,488 (17.1) & 90 (20.5)&\\
\quad Quintile 5 & 34,947 (13.5) & 72 (16.4)&\\
Year of first exacerbation (mean (SD)) & 11.25 (4.45) & 12.07 (4.30)& $\vecZ$\\
\hline
FEV1 \% predicted (mean (SD)) & 30.14 (31.24) & 27.23 (29.93) & $\vecX$\\
MRC (\%) &  & & $\vecX$ \\
\quad Grade 1 & 59,433 (22.9) & 75 (17.0) &\\
\quad Grade 2 & 109,283 (42.1) & 188 (42.7) &\\
\quad Grade 3 & 62,343 (24.0) & 108 ( 24.5)&\\
\quad Grade 4 & 24,377 (9.4) & 57 (13.0)&\\
\quad Grade 5 & 4,147 (1.6) & 12 (2.7)&\\
Latency days (mean (SD)) & 332.99 (429.28) & 318.26 (413.80)& $\vecZ$\\
Severity of first exacerbation  (\%) & 24,065 (9.3) & 52 (11.8)& $\vecX$\\
\hline
On oxygen therapy (\%) & 801 (0.3) & 4 (0.9)& $\vecX$\\
On statin (\%) & 27,328 (10.5) & 17 (3.9)& $\vecX$\\
Statin previous year (mean (SD)) & 1.47 (4.99) & 1.67 (5.50) & $\vecX$\\
On ICS (\%)& 101,856 (39.2) & 128 (29.1) & $\vecX$\\
ICS previous year (\%) & 58,140 (22.4) & 45 (10.2)& $\vecX$\\
On LAMA (\%) & 82,658 (31.8) & 92 ( 20.9) & $\vecX$\\
LAMA previous year (\%) & 44,030 (17.04) & 32 (7.3)& $\vecX$\\
On LABA (\%) & 98,797 (38.1) & 122 ( 27.7) & $\vecX$\\
LABA  previous year (\%) & 54,862 (21.1) & 42 (9.5) & $\vecX$\\
On other medications (\%) & 130,944 (50.4) & 440 (100.0) & $\vecZ$\\
\hline
Pneumonia LRTI (\%) & 89,027 (34.3) & 202 (45.9)& $\vecZ$\\
Active tuber (\%)& 680 (0.3) & 7 (1.6) & $\vecZ$\\
Nasal polyps (\%) & 2,856 (1.1) & 6 (1.4)& $\vecZ$\\
Heart failure (\%) & 7,071 (2.7) & 18 (4.1)& $\vecZ$\\
Asthma (\%) & 56,370 (21.7) & 133 (30.2)& $\vecZ$\\
Bronchiectasis (\%) & 3,466 (1.3) & 45 (10.2) & $\vecZ$\\
Abnormal heart rhythm (\%) & 1,317 (0.5) & 3 (0.7)& $\vecZ$\\
Ischaemic heart disease (\%) & 15,793 (6.1) & 37 (8.4)& $\vecZ$\\
Gastroesophageal reflux disease (\%) & 1,133 (0.4) & 1 (0.2)& $\vecZ$\\
Lung cancer (\%) & 2,008 (0.8) & 4 (0.9)& $\vecZ$\\
\bottomrule
\end{tabular}
\end{adjustbox}
\begin{tablenotes}[flushleft]
\footnotesize
\linespread{0.9}\selectfont
\item BMI: body-mass index. IMD$\_5$: index of Multiple Deprivation (Quintile 1 = most deprived; Quintile 5 = least deprived). COPD: chronic obstructive pulmonary disease. Year of first exacerbation: the number of years between 2024 and the year of first recorded exacerbation. FEV1: forced expiratory volume in one second (\% predicted). MRC: medical research council (Grade 1 = most mild limitation; Grade 5 = most severe limitation). Latency days: latency from COPD diagnosis to first exacerbation. ICS: inhaled corticosteroids.  LAMA: long-acting muscarinic receptor antagonist. LABA: long-acting $\beta$ agonist. LRTI: lower respiratory tract infection. 
\end{tablenotes}
\end{table}

\begin{figure}[h]
  \centering
   \includegraphics[scale=0.6]{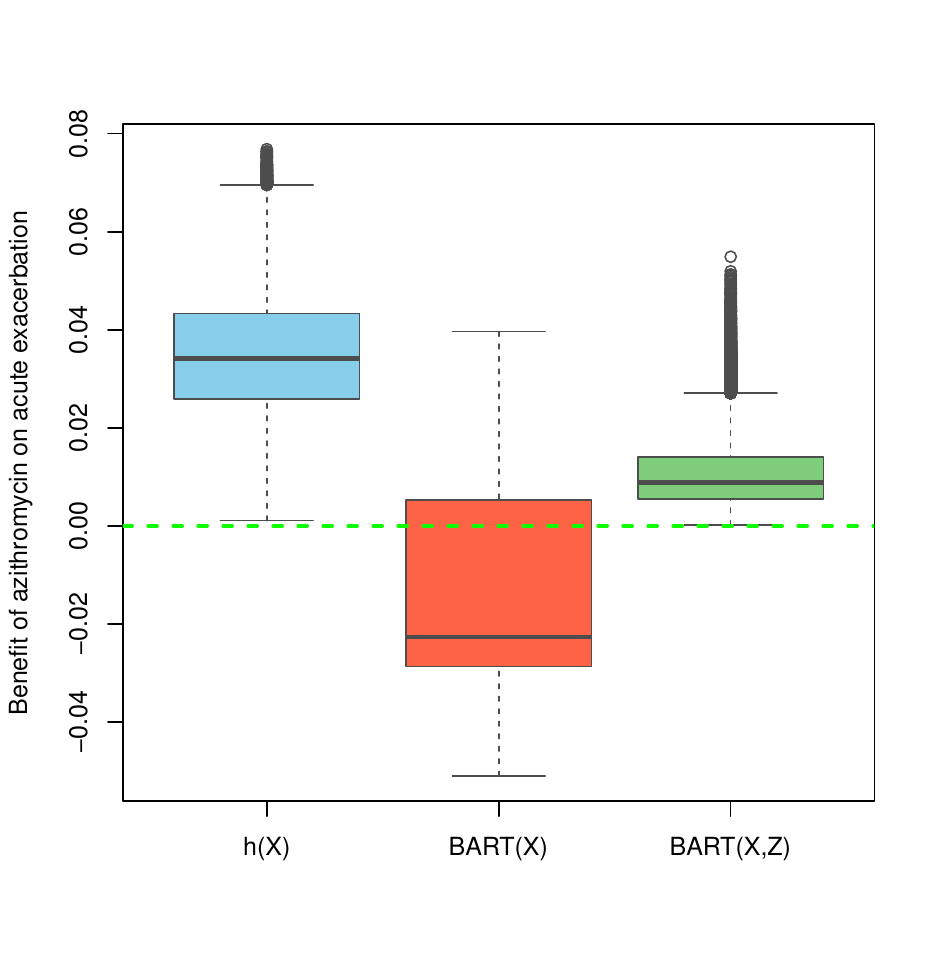}
    \caption{Across-patient distributions of treatment benefits predicted by $h(\vecx)$ and estimated by BART, where BART($\vecX$) indicates no adjustment for confounding, whereas BART($\vecX, \vecZ$) indicates adjustment for confounding.}
     \label{fig:6}
\end{figure}

We included 260,023 patients in the analysis: 440 with azithromycin initiation during the treatment window and 259,583 without initiation. 
Among those who did not initiate azithromycin, 35,842 (13.8\%) experienced the outcome; among those who did, 70 (15.9\%) experienced it.
The baseline characteristics are summarized in Table~\ref{tab:1}, including patient demographics,  COPD severity, COPD-related concomitant medications, and comorbidities.
Evaluating the TBP on every patient in the dataset yields the predicted benefits, shown as the blue boxplot in Figure~\ref{fig:6}. 
The predictions indicate a positive benefit of azithromycin on the outcome $Y$, with a mean of risk reduction of $0.0348$ and a standard deviation of $0.0119$.

Before evaluating the predictive performance of the TBP using the moderate calibration curve, we estimate $\tau(\vecx, \vecz)$ instead of $\tau_0(\vecx)$ using BART.
Ignoring $\vecZ$ in estimating $\mu_0(a, \vecx)$ yields a biased estimator of $\tau_0(\vecx)$, which can lead to misleading evaluation results.
To illustrate, we first estimate $\mu_0(a, \vecx)$ with default BART ($m = 100$ trees, $800$ burn-in iterations and $l = 500$ posterior draws) and derive the posterior distribution of $\mu_0(0, \vecx) - \mu_0(1, \vecx)$. 
For each observed $\vecx$, we take the posterior mean as the estimated value of $\tau_0(\vecx)$.
In Figure~\ref{fig:6}, the orange boxplot shows the across-patient distribution of estimated treatment benefits from BART without adjustment for $\vecZ$.
This impression of harm is likely due to uncontrolled confounding.
Therefore, we need to estimate $\mu(a, \vecx, \vecz)$ and derive the posterior distribution of $\tau(\vecx, \vecz)$. 
Under the same BART settings, we obtain $l =500$ posterior draws of $\tau$.
For each observed $(\vecx, \vecz)$, we take the posterior mean as the estimated value of $\tau(\vecx, \vecz)$.
The green boxplot in Figure~\ref{fig:6} shows the across-patient distribution of estimated treatment benefits from BART with adjustment for $\vecZ$.
This distribution is skewed to the right, with a median of $0.0088$ and an inter-quartile range of $0.0086$.
It shows a small average benefit of azithromycin initiation in preventing acute exacerbations or deaths across all patients in the cohort, which is $0.0104$.

\begin{figure}[h]
    \centering
    \includegraphics[scale=0.2]{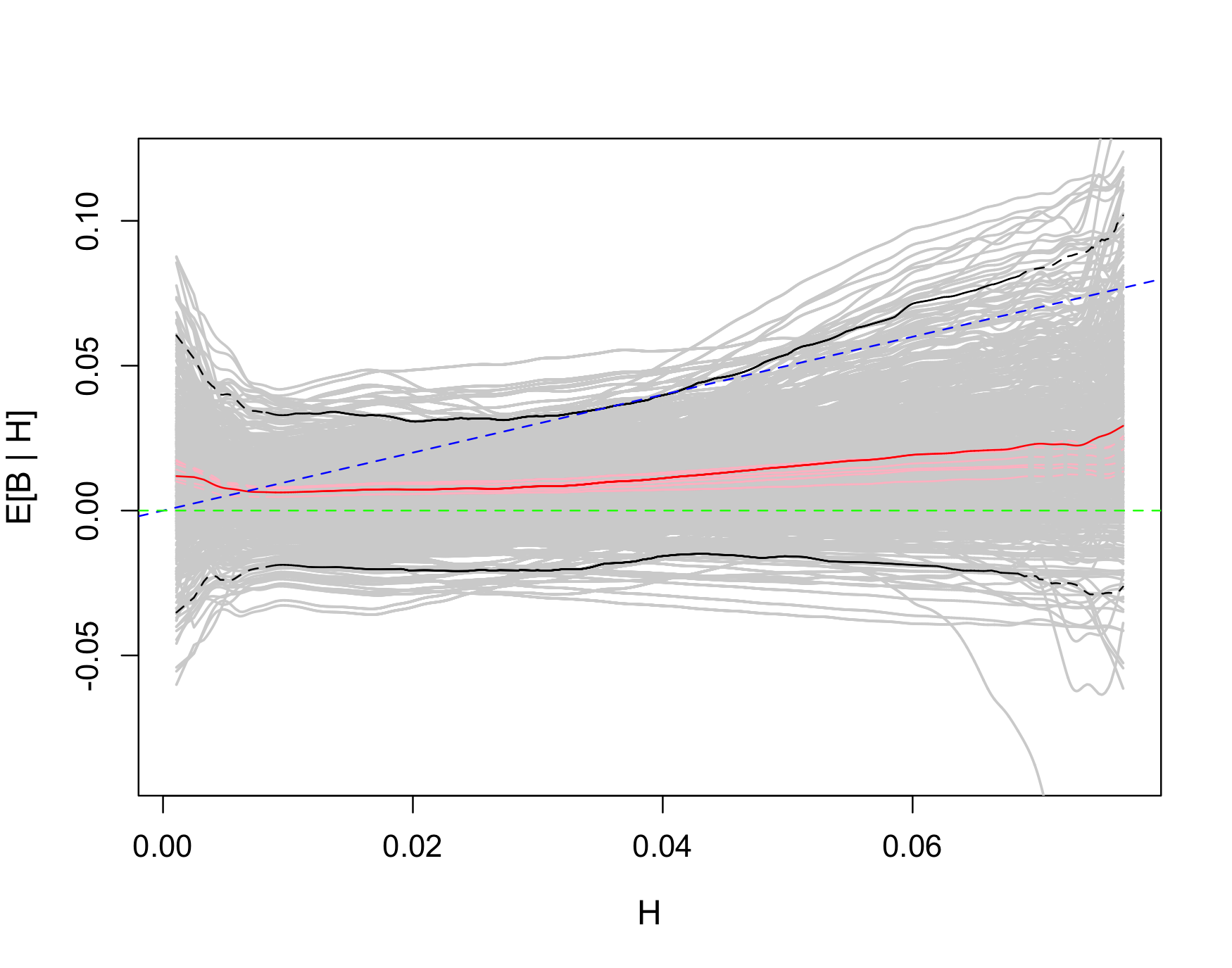}
    \caption{Estimated moderate calibration curves for the TBP. Gray curves represent 500 estimated calibration curves, the red curve is their mean, black dashed curves are 95\% quantile boundaries, the blue dash line is the reference line (perfect calibration), and pink dashed curves are posterior mean curves obtained from different starting points by varying the random seed.}
    \label{fig:7}
\end{figure}

We use these $500$ posterior draws of $\tau(\vecx,\vecz)$ to derive the $500$ posterior draws of the moderate calibration curve for $h(\vecx)$ using the estimator defined in Equation~(\ref{eq:cali_est_continous}). 
For every $\tau^{(l)}(\vecx, \vecz)$, a kernel regression smoother was used to estimate the $l$-th moderate calibration curve, using a Gaussian kernel and a bandwidth of $0.002$. 
These $500$ posterior draws of moderate calibration curves for $h(\vecx)$ are illustrated by the gray curves in Figure~\ref{fig:7}. 
For $H \in [0.005, 0.04]$, the moderate calibration curves show little dispersion; for $H < 0.005$ or $H > 0.04$, the variance is noticeably higher.
The red curve indicates the posterior mean; the green dashed line represents the 45-degree diagonal line, and the black dashed curves are the 95\% quantile bounds.
The moderate calibration plot shows that the constructed TBP is not moderately calibrated since the red curve is below the diagonal line and relatively flat. 
Therefore, TBP predicts a greater benefit of azithromycin for acute exacerbations or deaths than was detected with BART in the study cohort.

The pink dashed curves in Figure~\ref{fig:7} are posterior mean curves obtained from different starting points induced solely by varying the random seed. They show that the variation across starting points is smaller than the posterior variation under a single seed, which supports local stability. Further discussion of convergence is provided in Appendix E.

\section{Discussion} 
We demonstrated how to obtain the posterior distribution of the predictive performance measures for TBP as target parameters.
In particular, we obtained posterior draws of the function $\tau(\vecx,\vecz)$ from BART and then derived posterior draws of the measures: the $C_b$ index and the moderate calibration curve.
This approach relies on correct identification $\tau(\vecx,\vecz)$ under the stated assumptions and subsequent inference on the parameters. 
Simulation studies showed that the proposed estimators of moderate calibration curves and $C_b$s accurately capture the true calibration and discrimination performances of TBP, respectively, and that the posterior means are close to the targets for both discrete and continuous treatment benefit cases when treatment allocations are approximately balanced. 
Lastly, in a case study, we showcase the proposed methodology by evaluating a TBP for a simplified snapshot of the COPD disease and intervention trajectories.
The application to acute exacerbations among COPD patients illustrated that our framework enables a quantitative evaluation of predictive performance.
It also shows that the constructed TBP is not well suited to represent the benefit of azithromycin initiation in this study cohort.

\begin{figure}[h]
  \centering
   \includegraphics[scale=0.4]{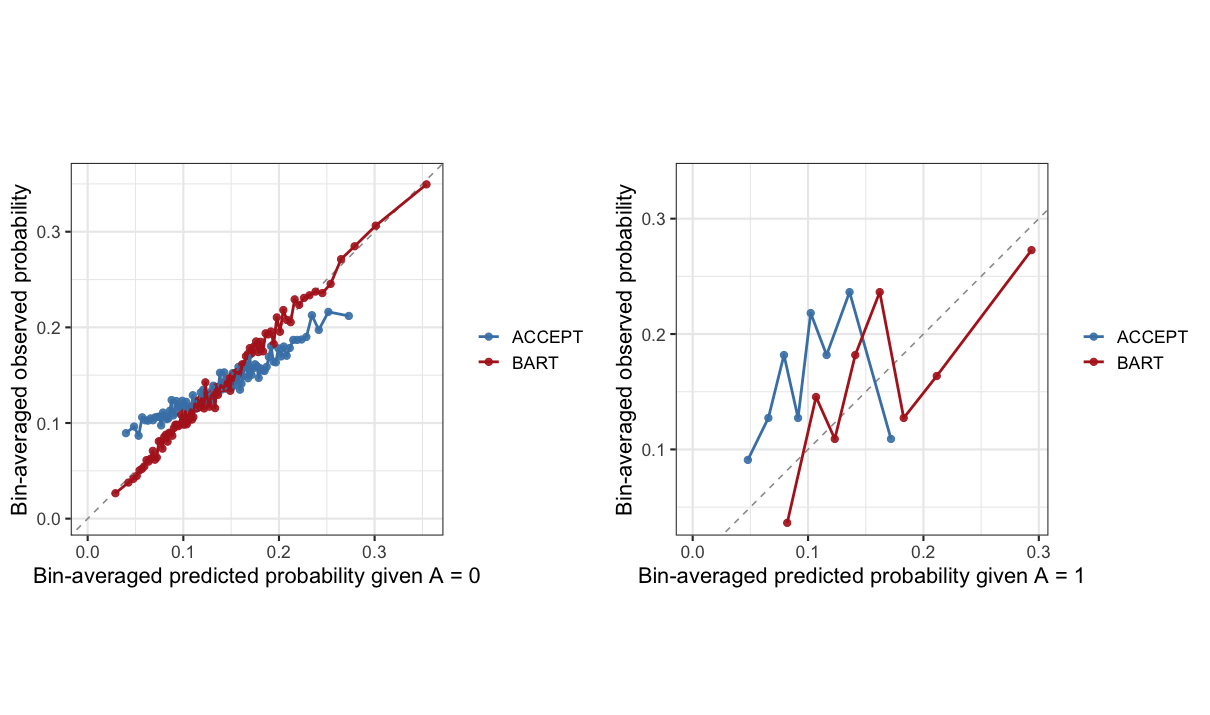}
    \caption{Observed calibration curves for ACCEPT and BART under two treatment arms. The untreated group is divided into 100 bins of 2,596 patients each, and the treated group into 8 bins of 55 patients each. }
     \label{fig:8}
\end{figure}

Why is this TBP not well suited to the estimand of interest in the target population?
We gain some insight by comparing how ACCEPT predicts risks $\mu_0(a, \vecx)$ and how BART estimates risks $\mu(a, \vecx, \vecz)$ within the corresponding treatment groups. 
Figure~\ref{fig:8} shows calibration plots of the predicted outcome risk using shared bins. 
The calibration bins for BART and ACCEPT are generated separately based on their corresponding values, so that the binning reflects the distribution of each method\textquotesingle s estimation or predictions.
For this study cohort, ACCEPT exhibits a slight tendency to overfit, as it stretches the lower end of the risk distribution and compresses the higher end in the untreated group. 
With the estimated hazard ratio, ACCEPT yields slightly lower risk estimates compared to the observed outcomes in the treated group.
This pattern might explain why the treatment benefits predicted by TBP tend to be greater than the treatment benefits estimated from BART.
However, there are far fewer patients who initiated azithromycin than those who did not; therefore, the data contain relatively little information on the treatment benefits in the target population.
As a result, any reasonable model is likely to perform less accurately.
This may further amplify the difference between the TBP\textquotesingle s predictions and BART\textquotesingle s estimates. 

From the standpoint of the study design, three key features help clarify this behavior.
First, recall that TBP is derived from three randomized controlled trials (RCTs) in which daily azithromycin was continued for 12 months, while our study cohort focuses on azithromycin initiation within one month.
Thus, the larger azithromycin benefit predicted by the TBP may be due to the longer exposure duration in the RCTs than in the study cohort. 
Second, as observational data typically involve greater levels of non-adherence and treatment switching, this intention-to-treat effect in the observational cohort is reasonably expected to be smaller than the effect reported in the RCTs.
Third, due to the limited number of treatment initiations and the observed second acute exacerbations, the target estimand is defined as a composite endpoint.
This endpoint differs from the outcome that the TBP was intended to predict, but it allowed evaluation in the observational data.
Consequently, the TBP developed using RCTs may not be well suited to predict the benefits of azithromycin initiation in the study cohort.

On the other hand, electronic health records are usually subject to large amounts of missing data.
In Section~\ref{sec:copd}, we addressed missing data in a separate imputation step before Bayesian nonparametric model estimation.
For computational simplicity, we used a single imputation instead of multiple imputations, so the uncertainty due to missing data was not fully accounted for.
Apart from this, our first intuition was that a Bayesian method should be able to handle missing data during iterative estimation, so it is possible to handle missingness within BART rather than requiring a separate step.
In this way, the posterior distribution of the measures would automatically reflect the uncertainty due to missing data.
In R, several BART packages are available on CRAN, some of which include built-in methods for handling missing data (\eg, the \pkg{BART} and \pkg{bartMachine} packages), while others do not (\eg, the \pkg{BayesTree} and \pkg{dbarts} packages).
In version 2.9.9 of the \pkg{BART} package, the \fun{pbart} function that we used does not handle missing data.
Another function, \fun{gbart}, addresses missing covariates by randomly copying covariate values from other records while ignoring the outcome, which can bias associations towards the null \cite{sparapani2021nonparametric}.
In contrast, the \pkg{bartMachine} package handles missing data using missingness patterns as information when it is predictive of the outcome \cite{kapelner2016bartmachine}.
Their approach is more closely aligned with the goal of handling missing data directly within the trees.
Overall, there is still much to explore regarding how BART handles missing data for treatment benefit estimation.

\pagebreak
\bibliographystyle{ama} 
\bibliography{biblio.bib}

@article{riley2024evaluation,
  title     = {Evaluation of clinical prediction models (part 2): how to undertake an external validation study},
  author    = {Riley, Richard D and Archer, Lucinda and Snell, Kym  and Ensor, Joie and Dhiman, Paula and Martin, Glen P and Bonnett, Laura J and Collins, Gary S},
  journal   = {BMJ},
  volume    = {384},
  year      = {2024},
  publisher = {British Medical Journal Publishing Group}
}

@article{spiegelhalter1986probabilistic,
  title={Probabilistic prediction in patient management and clinical trials},
  author={Spiegelhalter, David J},
  journal={Statistics in medicine},
  volume={5},
  number={5},
  pages={421--433},
  year={1986},
  publisher={Wiley Online Library}
}

@article{xia2025evaluating,
  title={Evaluating treatment benefit predictors using observational data: contending with identification and confounding bias},
  author={Xia, Yuan and Sadatsafavi, Mohsen and Gustafson, Paul},
  journal={American Journal of Epidemiology},
  pages={kwaf239},
  year={2025},
  publisher={Oxford University Press}
}

@Inbook{Kennedy2016,
author={Kennedy, Edward H.},
title={Semiparametric Theory and Empirical Processes in Causal Inference},
bookTitle={Statistical Causal Inferences and Their Applications in Public Health Research},
year={2016},
publisher={Springer International Publishing},
address={Cham},
pages={141--167},
isbn={978-3-319-41259-7},
doi={10.1007/978-3-319-41259-7_8},
}

@article{abrevaya2015estimating,
  title={Estimating conditional average treatment effects},
  author={Abrevaya, Jason and Hsu, Yu-Chin and Lieli, Robert P},
  journal={Journal of Business \& Economic Statistics},
  volume={33},
  number={4},
  pages={485--505},
  year={2015},
  publisher={Taylor \& Francis}
}

@article{robertson2021assessing,
  title={Assessing heterogeneity of treatment effects in observational studies},
  author={Robertson, Sarah E and Leith, Andrew and Schmid, Christopher H and Dahabreh, Issa J},
  journal={American Journal of Epidemiology},
  volume={190},
  number={6},
  pages={1088--1100},
  year={2021},
  publisher={Oxford University Press}
}

@article{chipman2010bart,
  title={BART: Bayesian additive regression trees},
  author={Chipman, Hugh A and George, Edward I and McCulloch, Robert E},
  journal={Annals of Applied Statistics},
  volume={6},
  number={1},
  pages={266--298},
  year={2010},
  publisher={Institute of Mathematical Statistics}
}

@article{greenland2017and,
  title={For and against methodologies: some perspectives on recent causal and statistical inference debates},
  author={Greenland, Sander},
  journal={European journal of epidemiology},
  volume={32},
  pages={3--20},
  year={2017},
  publisher={Springer}
}

@techreport{bierens1988nadaraya,
  title={The nadaraya-watson kernel regression function estimator},
  author={Bierens, Hermanus Josephus},
  year={1988},
  institution = {VU University Amsterdam, Faculty of Economics, Business Administration and Econometrics},
  address = {Amsterdam}
}

@article{foster2023orthogonal,
  title={Orthogonal statistical learning},
  author={Foster, Dylan J and Syrgkanis, Vasilis},
  journal={The Annals of Statistics},
  volume={51},
  number={3},
  pages={879--908},
  year={2023},
  publisher={Institute of Mathematical Statistics}
}

@article{calverley2023contemporary,
  title={Contemporary concise review 2022: chronic obstructive pulmonary disease},
  author={Calverley, Peter MA and Walker, Paul P},
  journal={Respirology},
  volume={28},
  number={5},
  pages={428--436},
  year={2023},
  publisher={Wiley Online Library}
}

@article{wolf2019data,
  title={Data resource profile: clinical practice research Datalink (CPRD) aurum},
  author={Wolf, Achim and Dedman, Daniel and Campbell, Jennifer and Booth, Helen and Lunn, Darren and Chapman, Jennifer and Myles, Puja},
  journal={International journal of epidemiology},
  volume={48},
  number={6},
  pages={1740--1740g},
  year={2019},
  publisher={Oxford University Press}
}

@article{adibi2020acute,
  title={The acute COPD exacerbation prediction tool (ACCEPT): a modelling study},
  author={Adibi, Amin and Sin, Don D and Safari, Abdollah and Johnson, Kate M and Aaron, Shawn D and FitzGerald, J Mark and Sadatsafavi, Mohsen},
  journal={The Lancet Respiratory Medicine},
  volume={8},
  number={10},
  pages={1013--1021},
  year={2020},
  publisher={Elsevier}
}

@article{olijnyk2022understanding,
  title={Understanding intent to treat analyses: an important lesson from the international cooperative study on the timing of aneurysm surgery},
  author={Olijnyk, L and Darsaut, TE and {\"O}hman, J and Raymond, J},
  journal={Neurochirurgie},
  volume={68},
  number={5},
  pages={471--473},
  year={2022},
  publisher={Elsevier}
}

@article{sparapani2021nonparametric,
  title={Nonparametric machine learning and efficient computation with Bayesian additive regression trees: The BART R package},
  author={Sparapani, Rodney and Spanbauer, Charles and McCulloch, Robert},
  journal={Journal of Statistical Software},
  volume={97},
  pages={1--66},
  year={2021}
}

@article{kapelner2016bartmachine,
  title={bartMachine: Machine learning with Bayesian additive regression trees},
  author={Kapelner, Adam and Bleich, Justin},
  journal={Journal of Statistical Software},
  volume={70},
  pages={1--40},
  year={2016}
}

@article{chipman1998bayesian,
  title={Bayesian CART model search},
  author={Chipman, Hugh A and George, Edward I and McCulloch, Robert E},
  journal={Journal of the American Statistical Association},
  volume={93},
  number={443},
  pages={935--948},
  year={1998},
  publisher={Taylor \& Francis}
}

@article{hill2011bayesian,
  title={Bayesian nonparametric modeling for causal inference},
  author={Hill, Jennifer L},
  journal={Journal of Computational and Graphical Statistics},
  volume={20},
  number={1},
  pages={217--240},
  year={2011},
  publisher={Taylor \& Francis}
}

@article{hahn2020bayesian,
  title={Bayesian regression tree models for causal inference: Regularization, confounding, and heterogeneous effects (with discussion)},
  author={Hahn, P Richard and Murray, Jared S and Carvalho, Carlos M},
  journal={Bayesian Analysis},
  volume={15},
  number={3},
  pages={965--1056},
  year={2020},
  publisher={International Society for Bayesian Analysis}
}

@article{murray2021log,
  title={Log-linear Bayesian additive regression trees for multinomial logistic and count regression models},
  author={Murray, Jared S},
  journal={Journal of the American Statistical Association},
  volume={116},
  number={534},
  pages={756--769},
  year={2021},
  publisher={Taylor \& Francis}
}

@article{albert2011azithromycin,
  title={Azithromycin for prevention of exacerbations of COPD},
  author={Albert, Richard K and Connett, John and Bailey, William C and Casaburi, Richard and Cooper Jr, J Allen D and Criner, Gerard J and Curtis, Jeffrey L and Dransfield, Mark T and Han, MeiLan K and Lazarus, Stephen C and others},
  journal={New England Journal of Medicine},
  volume={365},
  number={8},
  pages={689--698},
  year={2011},
  publisher={Mass Medical Soc}
}

@article{metlay2019diagnosis,
  title={Diagnosis and treatment of adults with community-acquired pneumonia. An official clinical practice guideline of the American Thoracic Society and Infectious Diseases Society of America},
  author={Metlay, Joshua P and Waterer, Grant W and Long, Ann C and Anzueto, Antonio and Brozek, Jan and Crothers, Kristina and Cooley, Laura A and Dean, Nathan C and Fine, Michael J and Flanders, Scott A and others},
  journal={American journal of respiratory and critical care medicine},
  volume={200},
  number={7},
  pages={e45--e67},
  year={2019},
  publisher={American Thoracic Society}
}

@article{whittaker2023frequency,
  title={Frequency and severity of respiratory infections prior to COPD diagnosis and risk of subsequent postdiagnosis COPD exacerbations and mortality: EXACOS-UK health care data study},
  author={Whittaker, Hannah and Nordon, Clementine and Rubino, Annalisa and Morris, Tamsin and Xu, Yang and De Nigris, Enrico and M{\"u}llerov{\'a}, Hana and Quint, Jennifer K},
  journal={Thorax},
  volume={78},
  number={8},
  pages={760--766},
  year={2023},
  publisher={BMJ Publishing Group Ltd}
}

@article{nahid2016official,
  title={Official American thoracic society/centers for disease control and prevention/infectious diseases society of America clinical practice guidelines: treatment of drug-susceptible tuberculosis},
  author={Nahid, Payam and Dorman, Susan E and Alipanah, Narges and Barry, Pennan M and Brozek, Jan L and Cattamanchi, Adithya and Chaisson, Lelia H and Chaisson, Richard E and Daley, Charles L and Grzemska, Malgosia and others},
  journal={Clinical infectious diseases},
  volume={63},
  number={7},
  pages={e147--e195},
  year={2016},
  publisher={Oxford University Press}
}

@article{wang2023impact,
  title={Impact of previous pulmonary tuberculosis on chronic obstructive pulmonary disease: baseline results from a prospective cohort study},
  author={Wang, Yide and Li, Zheng and Li, Fengsen},
  journal={Combinatorial Chemistry \& High Throughput Screening},
  volume={26},
  number={1},
  pages={93--102},
  year={2023},
  publisher={Bentham Science Publishers direct}
}

@article{fokkens2020_epos,
  author  = {Fokkens, Wytske J. and Lund, Valerie J. and Hopkins, Claire and Hellings, Peter W. and Kern, Robert and Reitsma, Selma and Toppila-Salmi, Sanna and others},
  title   = {European Position Paper on Rhinosinusitis and Nasal Polyps 2020 (EPOS2020)},
  journal = {Rhinology},
  year    = {2020},
  volume  = {58},
  number  = {Suppl 29},
  pages   = {1--464}
}

@article{andersson2023chronic,
author = {Andersson, Anders and Bergqvist, Joel and Lindberg, Anne and Zhou, Caddie and Nyberg, Fredrik and Hellgren, Johan and Stridsman, Caroline and Vanfleteren, Lowie E.G.W.},
title = {Chronic rhinosinusitis in COPD is associated with increased exacerbation risk},
volume = {62},
number = {suppl 67},
elocation-id = {PA3624},
year = {2023},
publisher = {European Respiratory Society},
journal = {European Respiratory Journal}
}

@misc{fda2013_azithro_qt,
  author  = {{U.S. Food and Drug Administration}},
  title   = {FDA Drug Safety Communication: Azithromycin (Zithromax or Zmax) and the risk of potentially fatal heart rhythms},
  year    = {2013},
  month   = {mar},
  howpublished = {Online}
}

@article{kim2021exploring,
  title={Exploring the impact of number and type of comorbidities on the risk of severe COPD exacerbations in Korean Population: a Nationwide Cohort Study},
  author={Kim, Youngmee and Kim, Ye-Jee and Kang, Yu Mi and Cho, Won-Kyung},
  journal={BMC pulmonary medicine},
  volume={21},
  number={1},
  pages={151},
  year={2021},
  publisher={Springer}
}

@article{gibson2017effect,
  title={Effect of azithromycin on asthma exacerbations and quality of life in adults with persistent uncontrolled asthma (AMAZES): a randomised, double-blind, placebo-controlled trial},
  author={Gibson, Peter G and Yang, Ian A and Upham, John W and Reynolds, Paul N and Hodge, Sandra and James, Alan L and Jenkins, Christine and Peters, Matthew J and Marks, Guy B and Baraket, Melissa and others},
  journal={The Lancet},
  volume={390},
  number={10095},
  pages={659--668},
  year={2017},
  publisher={Elsevier}
}

@article{hosseini2019global,
  title={Global prevalence of asthma-COPD overlap (ACO) in the general population: a systematic review and meta-analysis},
  author={Hosseini, Mostafa and Almasi-Hashiani, Amir and Sepidarkish, Mahdi and Maroufizadeh, Saman},
  journal={Respiratory research},
  volume={20},
  number={1},
  pages={229},
  year={2019},
  publisher={Springer}
}

@article{smith2020british,
  title={British Thoracic Society guideline for the use of long-term macrolides in adults with respiratory disease},
  author={Smith, David and Du Rand, Ingrid A and Addy, Charlotte and Collyns, Timothy and Hart, Simon and Mitchelmore, Philip and Rahman, Najib and Saggu, Ravijyot},
  journal={BMJ open respiratory research},
  volume={7},
  number={1},
  year={2020},
  publisher={British Thoracic Society}
}

@article{du2016bronchiectasis,
  title={Bronchiectasis as a comorbidity of chronic obstructive pulmonary disease: a systematic review and meta-analysis},
  author={Du, Qingxia and Jin, Jianmin and Liu, Xiaofang and Sun, Yongchang},
  journal={PloS one},
  volume={11},
  number={3},
  pages={e0150532},
  year={2016},
  publisher={Public Library of Science San Francisco, CA USA}
}

@article{schiefer2018current,
  title={Current perspective: osimertinib-induced QT prolongation: new drugs with new side-effects need careful patient monitoring},
  author={Schiefer, Mart and Hendriks, Lizza EL and Dinh, Trang and Lalji, Ulrich and Dingemans, Anne-Marie C},
  journal={European Journal of Cancer},
  volume={91},
  pages={92--98},
  year={2018},
  publisher={Elsevier}
}

@article{abdullah2019relationship,
  title={Relationship of atrial fibrillation to outcomes in patients hospitalized for chronic obstructive pulmonary disease exacerbation},
  author={Abdullah, Abdullah Sayied and Eigbire, George and Ali, Mohamed and Awadalla, Mohanad and Wahab, Abdul and Ibrahim, Hisham and Salama, Amr and Alweis, Richard},
  journal={Journal of atrial fibrillation},
  volume={12},
  number={2},
  pages={2117},
  year={2019}
}

@article{dunker2016impact,
  title={Impact of the FDA warning for azithromycin and risk for QT prolongation on utilization at an academic medical center},
  author={Dunker, Abby and Kolanczyk, Denise M and Maendel, Caitlin M and Patel, Amit R and Pettit, Natasha N},
  journal={Hospital Pharmacy},
  volume={51},
  number={10},
  pages={830--833},
  year={2016},
  publisher={SAGE Publications Sage CA: Los Angeles, CA}
}

@article{kunisaki2018exacerbations,
  title={Exacerbations of chronic obstructive pulmonary disease and cardiac events. A post hoc cohort analysis from the SUMMIT randomized clinical trial},
  author={Kunisaki, Ken M and Dransfield, Mark T and Anderson, Julie A and Brook, Robert D and Calverley, Peter MA and Celli, Bartolome R and Crim, Courtney and Hartley, Benjamin F and Martinez, Fernando J and Newby, David E and others},
  journal={American journal of respiratory and critical care medicine},
  volume={198},
  number={1},
  pages={51--57},
  year={2018},
  publisher={American Thoracic Society}
}

@misc{DerbyshireJAPC_Azithromycin_2022,
  author       = {{Derbyshire Joint Area Prescribing Committee}},
  title        = {Shared Care Agreement: Azithromycin for Use in Adult Respiratory Infections},
  institution  = {Derbyshire Medicines Management},
  year         = {2022},
  howpublished = {Online}
}

\section*{Appendix}

\subsection*{Appendix A. Bayesian additive regression trees}
This section reviews BART, especially for binary outcomes, and explains what it outputs, how those outputs are produced, and how they are used to estimate the $\tau(\vecx,\vecz)$.
BART, proposed by Chipman et al. (2010)\cite{chipman2010bart}, is a Bayesian nonparametric sum-of-trees model, which treats the target unknown function(s) as parameters and takes draws from the posterior distribution of the parameters. 
The posterior distribution is derived from the distribution of the sum of trees.
In BART, computational tractability is achieved through structural restrictions and priors. Each tree is a finite binary decision tree (every internal node has exactly two children). The number of trees in the ensemble is set to a fixed finite number. In addition, the BART prior favors small trees via depth-penalizing split probabilities, which concentrates posterior mass on shallow structures with few terminal nodes. Together, these elements allow efficient posterior simulation while retaining the flexibility of the sum-of-trees model.

\textbf{Why BART?} BART is an alternative to traditional linear or logistic regression models. It models the relationship between the outcome, treatment, and covariates more flexibly. 
It does not require researchers to choose which treatment-by-covariate interaction terms to include, and it is also flexible, allowing us to estimate the $\tau(\vecx,\vecz)$ either indirectly or directly by choosing different prior specifications. 
In particular, if the outcome model for continuous $Y_i$ is 
\[
Y_i = \mu(a_i,\vecx_i,\vecz_i) + \epsilon_i, \quad  \epsilon_i \overset{\iid}{\sim} \text{N}(0, \sigma^2),
\]
and $s = 1$, we can place a BART prior on a single unknown outcome function $\mu(a_i, \vecx_i,\vecz_i)$ and then derive $\tau(\vecx,\vecz)$ via
$\tau(\vecx_i,\vecz_i) = \mu(1,\vecx_i,\vecz_i) -  \mu(0,\vecx_i,\vecz_i)$ \cite{hill2011bayesian}.
Alternatively, instead of estimating a single function, we can have two separate potential-outcome functions, $\mu^{(0)}(\vecx_i,\vecz_i)$ and $\mu^{(1)}(\vecx_i,\vecz_i)$, for non-treatment and treatment conditions, respectively.
Since $\tau(\vecx_i,\vecz_i) = \mu^{(1)}(\vecx_i,\vecz_i) - \mu^{(0)}(\vecx_i,\vecz_i)$, a common reparameterization for the outcome model is 
\[
Y_i  = \mu^{(0)}(\vecx_i,\vecz_i) - \tau(\vecx_i,\vecz_i)a_i + \epsilon_i.
\]
Consequently, BART priors can be placed on two unknown functions $\mu^{(0)}(\vecx_i,\vecz_i)$ and $\tau(\vecx_i,\vecz_i)$. There are also modified outcome models built on these two fundamental structures \cite{hahn2020bayesian}. In what follows, we focus on the indirect method that first estimates $\mu(a_i, \vecx_i,\vecz_i)$ from BART and then derives $\tau(\vecx_i,\vecz_i)$.

For binary outcome $Y_i$, Chipman et al. (2010) extended the development of BART for continuous outcome using a probit latent variable model \cite{chipman2010bart}:
\begin{align*}
    Y^*_i &= g(a_i, \vecx_i,\vecz_i) +  \varepsilon_i, \quad \varepsilon_i \overset{\iid}{\sim} \text{N}(0, 1),\\
    Y_i &= \I(Y^*_i > 0),
\end{align*}
where $\I(\cdot)$ is the indicator function, and $Y^*_i$ is an augmented latent variable with unit variance. Then, all $Y_i$ are independent Bernoulli random variables with conditional mean $\mu(a_i, \vecx_i,\vecz_i) = \Phi(g(a_i, \vecx_i,\vecz_i))$, where $\Phi(\cdot)$ is the standard normal CDF.  
The probit latent-variable model is not the unique approach to a binary outcome in the BART framework. 
A logit model has also been used in BART via Pólya–Gamma augmentation \cite{murray2021log}.

BART models the function $g$ as a sum of trees,
\[g(a_i, \vecx_i,\vecz_i) = \sum^m_{j=1} g_j(a_i, \vecx_i,\vecz_i; \vecT_j, \vecM_j), 
\]
where $m$ is a fixed finite number of trees and each component $g_j$ is a tree-based function. The $j$-th tree-based function $g_j$ maps an input $(a_i, \vecx_i,\vecz_i)$ to a leaf value, with the destination determined entirely by its structure $\vecT_j$ (splitting rules) and terminal-node parameters $\vecM_j$ (leaf values). Specifically, $g_j(a_i, \vecx_i,\vecz_i; \vecT_j, \vecM_j) = \nu_{jk} \in \vecM_{j}$, where $k$ denotes the index of the destined leaf in the $j$-th tree. Thus, $g$ is the deterministic map implied by the parameters $\{(\vecT_j, \vecM_j)\}^m_{j=1}$, and the initial uncertainty about $g$ is induced by the prior on these tree parameters.

The BART prior makes each tree a weak learner. 
In particular, for the $j$-th tree, the prior on $\vecT_j$ controls where and how often a node splits via a depth-penalized splitting probability 
$\alpha(1 + d)^{-\beta}$,
with defaults $\alpha = 0.95$ and $\beta = 2$, which favors shallow trees. Conditional on splitting, the splitting variable is chosen uniformly from the feasible variables. 
Then, given the selected variable, the splitting value is chosen uniformly from its admissible values at the node. Given a structure $\vecT_j$, the prior on each $\vecM_j$ governs the value returned for each $g_j$ map.
Consequently, a common leaf prior is a strongly shrinking normal prior on each leaf value $\nu_{jk} \overset{\iid}{\sim} \text{N}(0, \sigma_{\nu}^2)$ with small $\sigma_{\nu}^2$, which regularizes how much one tree contributes.

BART uses a Metropolis-within-Gibbs scheme: a Gibbs sampler on the outside and a Metropolis–Hastings algorithm (MH) step inside. Specifically, each iteration of the Gibbs sampler updates one tree $(\vecT_j, \vecM_j)$ at a time, conditional on all other trees and $\vecY^* = (Y_1^*, \cdots, Y_n^*)$.  After cycling through the $m$ trees, each $Y^*_i$ is updated from truncated normal distributions,
\begin{align*}
    Y^*_i \mid Y_i = 0 &\sim \text{N}\left(g(a_i,\vecx_i,\vecz_i),1\right) ~\text{truncated to}~ (-\infty, 0 ] ,\\
    Y^*_i \mid Y_i = 1 &\sim \text{N}\left(g(a_i,\vecx_i,\vecz_i),1\right) ~\text{truncated to}~  (0, \infty).
\end{align*}
In the tree-specific update, $\vecT_j$ is updated using MH, which is proposed given the current tree using one of four moves: grow a terminal node, prune a pair of terminal nodes, swap splitting rules of two interior nodes, or change the splitting rule of an interior node \cite{chipman1998bayesian, chipman2010bart}.
Because the model is additive, the data and $\vecY^*$ enter the update for tree $j$ only through the current partial residuals that are not explained by the other trees,
\begin{align*}
    \vecR_{j} &= \vecY^* - \sum_{j' \neq j} g_{j'}(\D; \vecT_{j'}, \vecM_{j'}),\\
    \vecR_{j} &\mid (\vecT_{j}, \vecM_{j}) \sim N\left(g_{j}(\D; \vecT_{j}, \vecM_{j}), \II_n \right),
\end{align*}
where $\vecR_j$ denotes a vector of partial residuals for all observations and $\II_n$ is the identity matrix.
As $\vecR_{j}$ conditional on $(\vecT_{j}, \vecM_{j})$ is multivariate normal, the collapsed marginal likelihood $f(\vecR_{j} \mid \vecT_j)$ for MH is available in closed form by integrating out $\vecM_j$. 
Conditional on the updated $\vecT_j$ and current $\vecR_{j}$, $\vecM_j$ is updated from $f(\vecM_j \mid \vecT_j, \vecR_{j})$.
Since $\vecR_{j} \mid (\vecT_{j}, \vecM_{j})$ is multivariate normal and $\vecM_j \mid \vecT_j$ follows a conjugate multivariate normal prior, the full conditional $f(\vecM_j \mid \vecT_j, \vecR_{j})$ is also multivariate normal. Consequently, by applying the backfitting to trees, \ie updating tree $j$ given the partial residual $\vecR_j$, the MH update for the structure $\vecT_j$ is computationally efficient and numerically stable, and conditional on $\vecT_j$, sampling $\vecM_j$ reduces to a simple Gibbs draw. 
Each Gibbs iteration yields a sum-of-trees evaluated in the $n$ observations. After a prespecified burn-in, we retain $l$ iterations, yielding an autocorrelated sample from the posterior of $g$.

\subsection*{Appendix B. Construction of the TBP to be evaluated}

\textbf{Hazard Ratio for First Exacerbation.} Albert et al. (2011) \cite{albert2011azithromycin} conducted a RCT of 1142 North American patients with COPD to assess whether adding daily oral azithromycin to usual care for one year reduces the frequency of acute exacerbations between March 2006 and June 2010. 
Of the patients, 570 were assigned to receive azithromycin (250 mg) and 572 to receive a placebo. They assumed a constant treatment hazard ratio over time and fit a Cox proportional-hazards model. The estimated hazard ratio for the time to the first exacerbation was 0.73, indicating a $27\%$ lower hazard with azithromycin.

The inclusion criteria for our study cohort are similar to those of the RCT, although not identical, because some measures are unavailable and because of sample size considerations. Both require COPD diagnosis, age $\geq 40$ years, smoking history, and $\text{FEV}_1/\text{FVC} < 0.7$. 
However, although our inclusion and exclusion criteria are based on those of the trial, its stricter requirements cannot be fully replicated in the observational data.
For instance, the trial used stricter criteria: $\ge 10$ pack-years of smoking; either continuous supplemental oxygen use or systemic glucocorticoids in the prior year; hospitalization for an acute COPD exacerbation; and absence of asthma. Moreover, the RCT evaluated daily azithromycin for 12 months, whereas our observational analysis focuses on initiation within one month. We therefore expect attenuation toward the null. The true hazard ratio in our study cohort is likely to be higher (closer to 1.0) than 0.73, reflecting the shorter exposure period. Therefore, we adapt the finding from the RCT to construct a TBP for azithromycin on exacerbations as an illustrative approximation to convey the general idea.

Let $\lambda(t\mid A, \vecX)$ denote the hazard function at time $t$ conditional on dichotomized treatment $A$ and covariates $\vecX$; let $T$ be the random time to exacerbation and $t$ a specific time point. Assume that at any given time $t$ among those who have not yet had an exacerbation, the hazard for azithromycin is $27\%$ lower than for placebo, conditional on the set $X$:
\[
\lambda(t \mid A = 1, \vecX = \vecx) = 0.73 \cdot \lambda(t \mid A = 0, \vecX = \vecx).
\]
Along with the hazard function, the cumulative hazard function is denoted as $\Lambda(t) = \int^t_0 \lambda(u)du$.
Under the proportional hazards assumption with a constant hazard ratio, the survival functions satisfy:
\[
S(t \mid A = 1,  \vecX = \vecx) = S(t \mid A = 0,  \vecX = \vecx)^{0.73},
\]
where $S(t \mid A, \vecX) = \P(T > t \mid A, \vecX = \vecx)$ represents the survival function conditional on treatment and covariates. This follows because cumulative hazard functions satisfy $\Lambda(t \mid A = 1, \vecX) = 0.73 \Lambda(t \mid A = 0, \vecX = \vecx)$ and $S(t \mid A, \vecX = \vecx) = \exp\{\Lambda(t \mid A, \vecX = \vecx)\}$. Therefore, given the probability that the exacerbation time exceeds $t$ without azithromycin ($A=0$) and covariates $\vecX$, we can predict the corresponding probability with azithromycin ($A=1$) for the same $X$.

\textbf{Exacerbation rate prediction tool}
Adibi et al. (2020) \cite{adibi2020acute} proposed a generalizable algorithm, called ACCEPT, to predict the individualized risk of acute COPD exacerbations over a 12 months. The algorithm was trained on 2,380 COPD patients pooled from three RCTs, including the trial analyzed by Albert et al. (2011)\cite{albert2011azithromycin}. 
Given baseline characteristics (exacerbation history in the last year, age, sex, BMI, smoking status, domiciliary oxygen therapy, lung function, symptom burden, and prior-year and current medication use), the algorithm produces individualized predictions of exacerbation risk over 12 months. 
For each patient, the algorithm models the time to each exacerbation using an accelerated failure time (AFT) model with a random intercept and the probability that an exacerbation is severe using a logistic regression model with another random intercept.
These two random effects follow a correlated bivariate normal distribution to capture that patients who exacerbate more frequently also tend to have more severe exacerbations.
For more details on the model specification, see Section 2 of the Supplementary Material in Adibi et al. (2020)\cite{adibi2020acute}.

We highlight a few important features here.
For $i$-th patient, the time to the $k$-th exacerbation is modeled using 
\begin{align*}
    \log \left(T_{ik}\right) = - u_i - \vecx_{-i}^T\beta + \varepsilon_{ik},
\end{align*}
where $T_{ik}$ is the time to the $k$-th exacerbation, $u_i$ is a patient-specific random effect, $\beta$ is a fixed effect vector, and $ \varepsilon_{ik}$ is random noise for this exacerbation.
Note that $\vecx_{-}$ is a vector of covariates that contains all mentioned patient characteristics except the history of exacerbations.
This model can be re-expressed in terms of hazard functions:
\begin{align*}
    \lambda(t_{ik} \mid A = 0, \vecX_{-} = \vecx_{-i}) &= \lambda_0\left(t_{ik} \cdot \theta(\vecX_{-}) \mid A = 0, \vecX_{-} = \vecx_{-i}\right)\cdot\theta(\vecx_{-i}),\\
    \theta(\vecx_{-i}) &= \exp(u_i + \vecx_{-i}^T\beta).
\end{align*}
In other words, in this AFT model, the true frailty $u_i$ is assumed to be constant over time, so the hazard for each exacerbation of patient $i$ is determined by the same $u_i$.
However, different patients have different $u_i$, and the model implies that patients who experience more exacerbations are more likely to have a higher underlying frailty.

Suppose that the full exacerbation history of $i$-th patient is denoted by $\mathcal{H}_i = \{(T_{ik}, Y_{ik}): k = 1,\cdots,n_i\}$, where $Y_{ik}$ is the severity of $k$-th exacerbation. 
The joint AFT-logistic model considers two latent patient-level random effects: $u_i$ and $v_i$.
After estimating all fixed effects and variance parameters of the joint model, individual predictions cannot be obtained simply by plugging in these estimates because all random effects remain unobserved.
For a new patient $i'$, ACCEPT computes the posterior distribution of $u_{i'}$ and $v_{i'}$ given $\mathcal{H}_{i'}$ and $\vecX_{-i'}$ ($\vecX = (\mathcal{H}, \vecX_{-})$), where both likelihood and prior depend on the plugged-in estimates of the model parameters.
Finally, we can obtain the posterior mean hazard by integrating the hazard function for patient $i'$ with respect to the posterior of $u_{i'}$ and $v_{i'}$.
This posterior mean usually does not have a closed form.
It can be approximated by drawing samples from the joint posterior distribution of $u_{i'}$ and $v_{i'}$, evaluating the hazard function at each draw, and then taking the average.

With the predicted hazards for the first exacerbation from ACCEPT algorithm, we can compute the cumulative hazards up to 12 months, denoted as $\hat{\Lambda}(12 \mid A=0, \vecX = \vecx)$. 
Then, the predicted probability of no exacerbation in 12 months is
\begin{align*}
    S(12 \mid A=0, \vecX = \vecx) = \exp(-\hat{\Lambda}(12 \mid A=0, \vecX = \vecx)).
\end{align*}
The predicted probability of having at least one exacerbation in 12 months is 
\begin{align*}
     1 - S(12 \mid A=0, \vecX = \vecx) = 1 - \exp(-\hat{\Lambda}(12 \mid A=0, \vecX = \vecx)),
\end{align*}
which we denoted as $\pi_{12}(\vecx)$.

\textbf{Treatment benefit predictor construction.}
Next, we stitch findings from both studies to discuss a TBP of the azithromycin benefit on exacerbations based on baseline patient characteristics.
Since $Y$ is defined as the indicator of having the event in two months and $s = 0$, we have $Y:= \I(T \leq 2)$ and can express $\tau_0(\vecx)$ as
\begin{align*}
    \text{E}(B \mid \vecX = \vecx) &= \P(Y = 1 \mid A = 0, \vecX = \vecx) - \P(Y = 1 \mid A = 1, \vecX = \vecx)\\
    &= \P(T \leq 2 \mid A = 0, \vecX = \vecx) - \P(T \leq 2 \mid A = 1, \vecX = \vecx)\\
        &= (1 - S(2 \mid A = 0, \vecX = \vecx)) - (1 - S(2 \mid A = 1, \vecX = \vecx))\\
    &= S(2 \mid A = 1, \vecX = \vecx) - S(2 \mid A = 0, \vecX = \vecx).
\end{align*}
Subsequently, we construct a TBP for azithromycin on acute exacerbation based on this relation by estimating
\begin{align*}
    S(2\mid A=0, \vecX = \vecx) &\approx S(12\mid A=0, \vecX = \vecx)^{1/6},\\
    S(2\mid A=1, \vecX = \vecx) &\approx S(2\mid A=0, \vecX = \vecx)^{0.73} \\
    &= \left(S(12\mid A=0, \vecX = \vecx)^{1/6}\right)^{0.73},
\end{align*}
where the second approximation further assumes an approximately constant hazard over the 12-month period. Denoting the predicted probability of at least one exacerbation over 12 months from ACCEPT as $\pi_{12}(\vecx)$, we can express the TBP as
\begin{align*}
    h(\vecx) &= \left(S(12\mid A=0, \vecX = \vecx)^{1/6}\right)^{0.73} - S(12\mid A=0, \vecX = \vecx)^{1/6}\\
    &= \left(\left(1 - \pi_{12}(\vecx)\right)^{1/6}\right)^{0.73} - \left(1 - \pi_{12}(\vecx)\right)^{1/6}.
\end{align*}

\subsection*{Appendix C. Selection of baseline confounders}
Confounders are variables that influence both initiating azithromycin and exacerbation. Major confounders typically arise from historical and current medication use, comorbidities, COPD severity, healthcare utilization, patient demographics, and calendar time. In addition to the variables in $\vecX$, we include the following additional variables to adjust for confounding.

In line with the Joint Area Prescribing Committee’s azithromycin guideline, azithromycin is considered for COPD patients after best-practice COPD care has been implemented \cite{DerbyshireJAPC_Azithromycin_2022}. Therefore, in addition to the medications used in ACCEPT, we adjust for relevant concomitant medications measured at baseline as confounders, including short-acting $\beta_2$-agonists (SABA), short-acting muscarinic antagonists (SAMA), fixed-dose SABA/SAMA combinations, macrolides, mucolytics, phosphodiesterase-4 (PDE4) inhibitors, and theophylline.

\begin{table}[th]
\caption{Comorbidities that may influence treatment and outcome.}
\label{tab:comorbidities}

\centering
\small
\setlength{\tabcolsep}{4pt}
\renewcommand{\arraystretch}{0.5}
\begin{adjustbox}{width=0.8\linewidth}
\begin{tabular}{lcc}
\toprule
Comorbidities & Azithromycin &  Acute exacerbation\\
 \midrule
 Pneumonia LRTI & Metlay et al. (2019)\cite{metlay2019diagnosis} & Whittaker et al. (2023)\cite{whittaker2023frequency}
\\
Active tuber & Nahid et al. (2016)\cite{nahid2016official} & Wang et al. (2023)\cite{wang2023impact}\\
Nasal polyps & Fokkens et al. (2020)\cite{fokkens2020_epos} & Andersson et al. (2023) \cite{andersson2023chronic}\\
Heart failure & U.S. Food and Drug Administration (2013)\cite{fda2013_azithro_qt}& Kim et al. (2021) \cite{kim2021exploring}\\
Asthma & Gubson et al. (2017)\cite{gibson2017effect}& Hosseini et al. (2019)\cite{hosseini2019global}\\
Bronchiectasis & Smith et al. (2020) \cite{smith2020british} & Du et al. (2016) \cite{du2016bronchiectasis}\\
Lung cancer & Schiefer et al. (2018) \cite{schiefer2018current} & Kim et al. (2021) \cite{kim2021exploring}\\
Arrhythmia & Smith et al. (2020) \cite{smith2020british}  & Abdullah et al. (2019)\cite{abdullah2019relationship}\\
Ischemuic heart disease & Dunker et al. (2016)\cite{dunker2016impact} & Kunisaki et al. (2018)\cite{kunisaki2018exacerbations}\\
\bottomrule
\end{tabular}
\end{adjustbox}
\end{table}

Another source of confounding arises from comorbidities that influence both the probability of initiating azithromycin and the risk of exacerbations. We summarize these comorbidities, with supporting evidence, in Table~\ref{tab:comorbidities}. The listed conditions are major precipitants or predictors of acute exacerbations. Among these comorbidities, some favor the consideration of azithromycin, whereas others may discourage its use due to safety concerns.

In addition, we include socioeconomic status, the interval between the COPD diagnosis date and the first exacerbation, and the scaled calendar year of the first exacerbation as confounders to capture healthcare access/utilization, disease severity, and temporal prescribing patterns that may influence azithromycin initiation and exacerbation risk.

\subsection*{Appendix D. COPD data, missingness, and checks}
This section profiles data completeness and variable distributions in the COPD cohort. The cohort contains 261,525 observations and 41 patient characteristics, including calendar time, 19 comorbidities, 2 variables that reflect COPD severity, 6 medications used in the past year, 7 current medications and 7 others. Breathlessness was assessed using the MRC, a 5-point ordinal measure ranging from 1 (`breathless only with strenuous exercise') to 5 (`too breathless to leave the house or breathless when dressing'). Lung function was measured using $\text{FEV}_1$, expressed as percent predicted (\% predicted) with values ranging from 0.19 to 99.72, where values below 10 reflect very severe impairment. BMI was calculated as weight in kilograms divided by height in meters squared, with values in the dataset ranging from 12.02 to 97.90. Socioeconomic status was assessed using the $\text{IMD}_5$, with 1 representing the least deprived and 5 the most deprived. Among the 19 comorbidities included, most were related to the cardiovascular, metabolic, or respiratory systems. Medications mainly recorded included inhaled corticosteroids (ICS), long-acting $\beta_2$-agonists (LABA), long-acting muscarinic antagonists (LAMA), oxygen therapy, and statins, with azithromycin considered the active treatment.

\begin{figure}
    \centering
    \includegraphics[scale = 0.7]{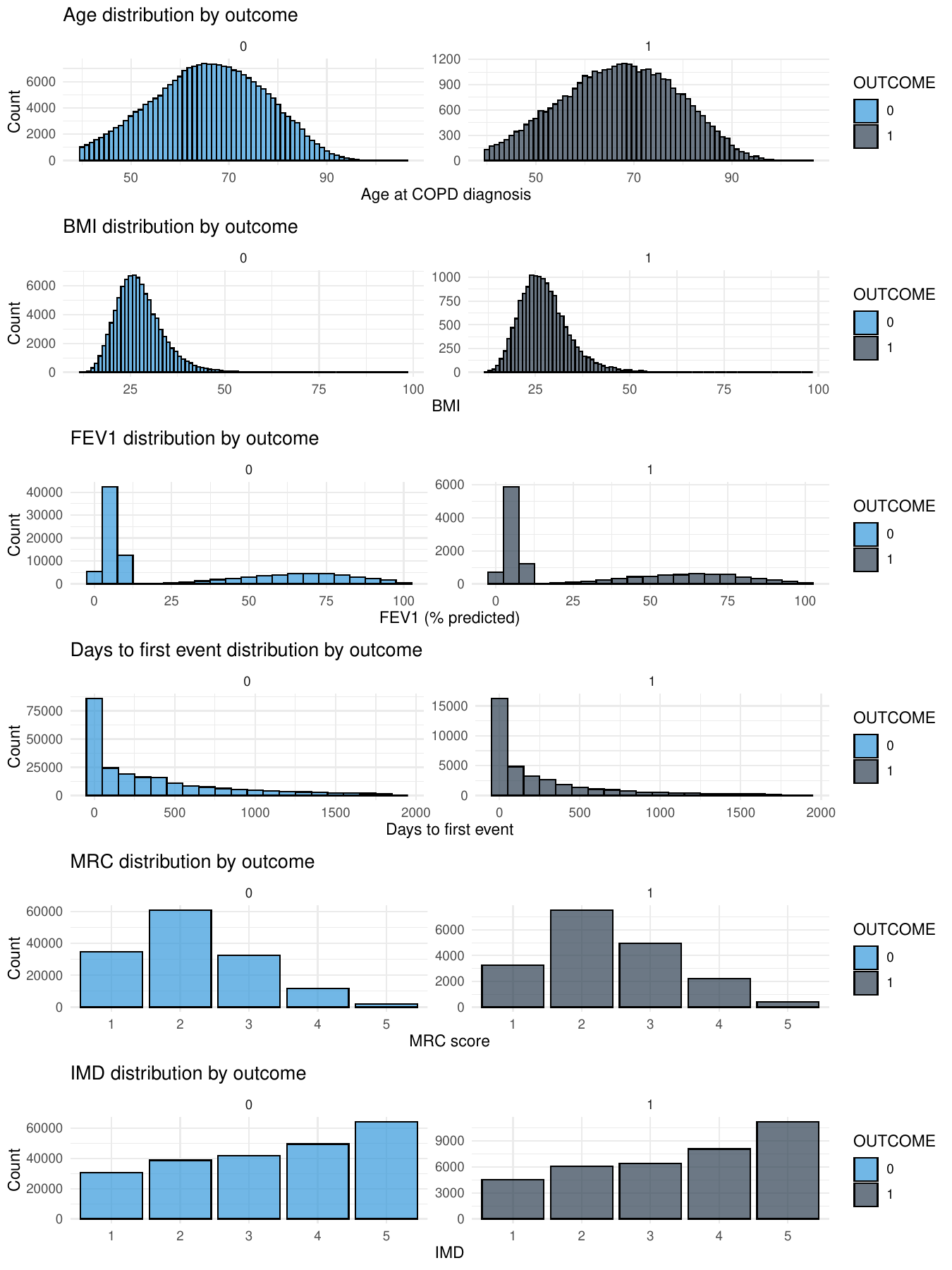}
    \caption{Histograms for selected covariates by outcome}
    \label{fig:orginal_distributions1}
\end{figure}

\begin{figure}
    \centering
    \includegraphics[scale = 0.7]{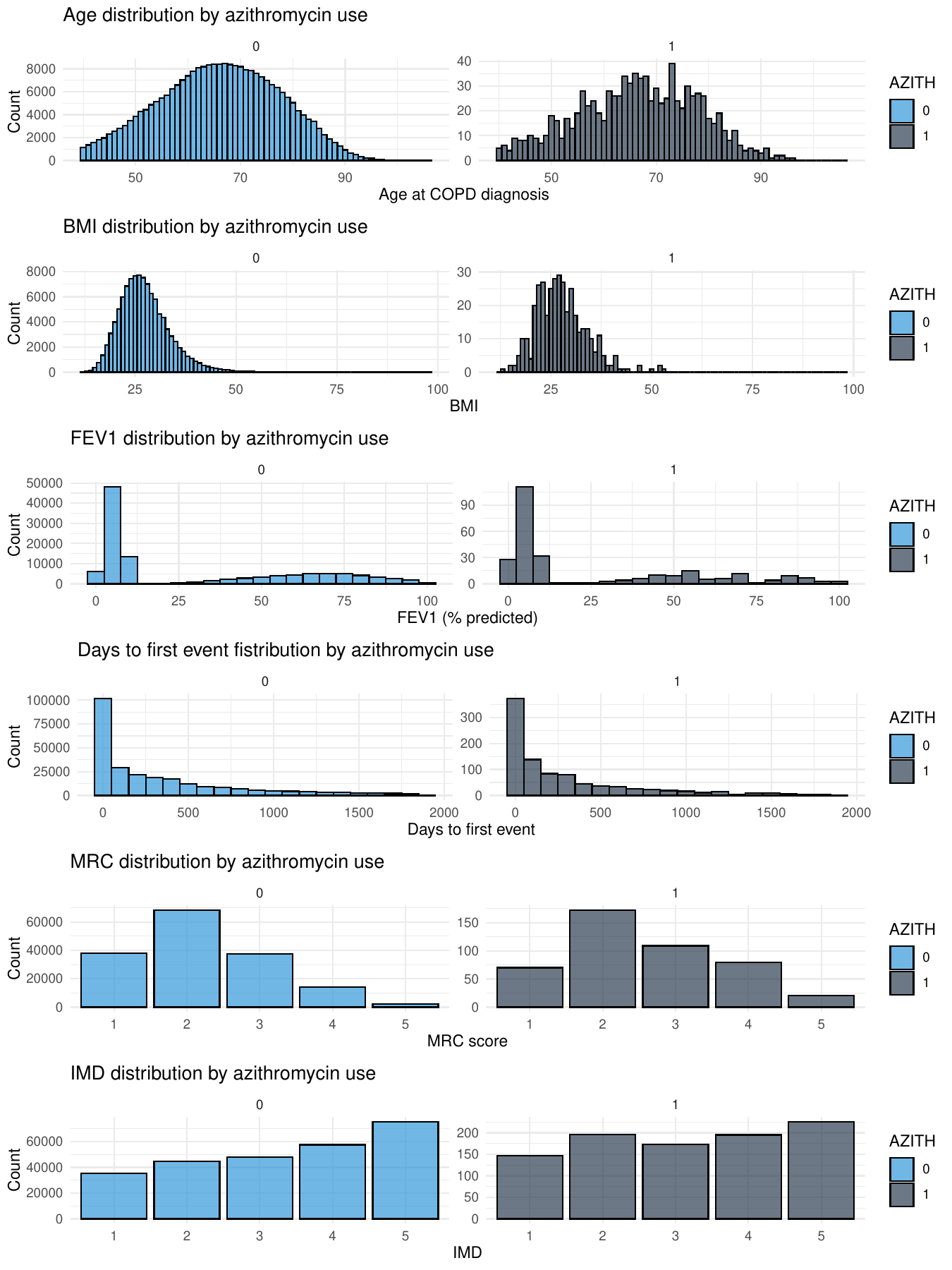}
    \caption{Histograms for selected covariates by the active treatment}
    \label{fig:orginal_distributions2}
\end{figure}

\textbf{Explore the relationships between variables.} 
We plot the distributions of some key variables, stratified by outcome and, separately, by active treatment (Figure~\ref{fig:orginal_distributions1} and Figure~\ref{fig:orginal_distributions2}). Visual inspection does not reveal clear differences between strata. The sample is notably imbalanced: there are far fewer observations with severe outcomes than without, and far fewer who received the active treatment of interest. 
The pairwise Pearson correlations among continuous variables (age, BMI, $\text{FEV}_1$, total number of previous statin prescriptions, days from COPD diagnosis to the first exacerbation) are near zero ($|r| \leq 0.10$). The greatest magnitudes are BMI \text{--} $\text{FEV}_1$ $(r = 0.10)$, age \text{--}  total number of previous statin prescriptions $(r = 0.097)$, BMI \text{--} total number of previous statin prescriptions $(r = 0.073)$, and age \text{--} BMI $(r = -0.070)$; the correlations with days from COPD diagnosis to the first exacerbation are $\leq 0.028$ in absolute value. These results indicate no strong pairwise linear associations and no concerning collinearity for multivariable modeling.

\begin{figure}
    \centering
    \includegraphics[width=1\linewidth]{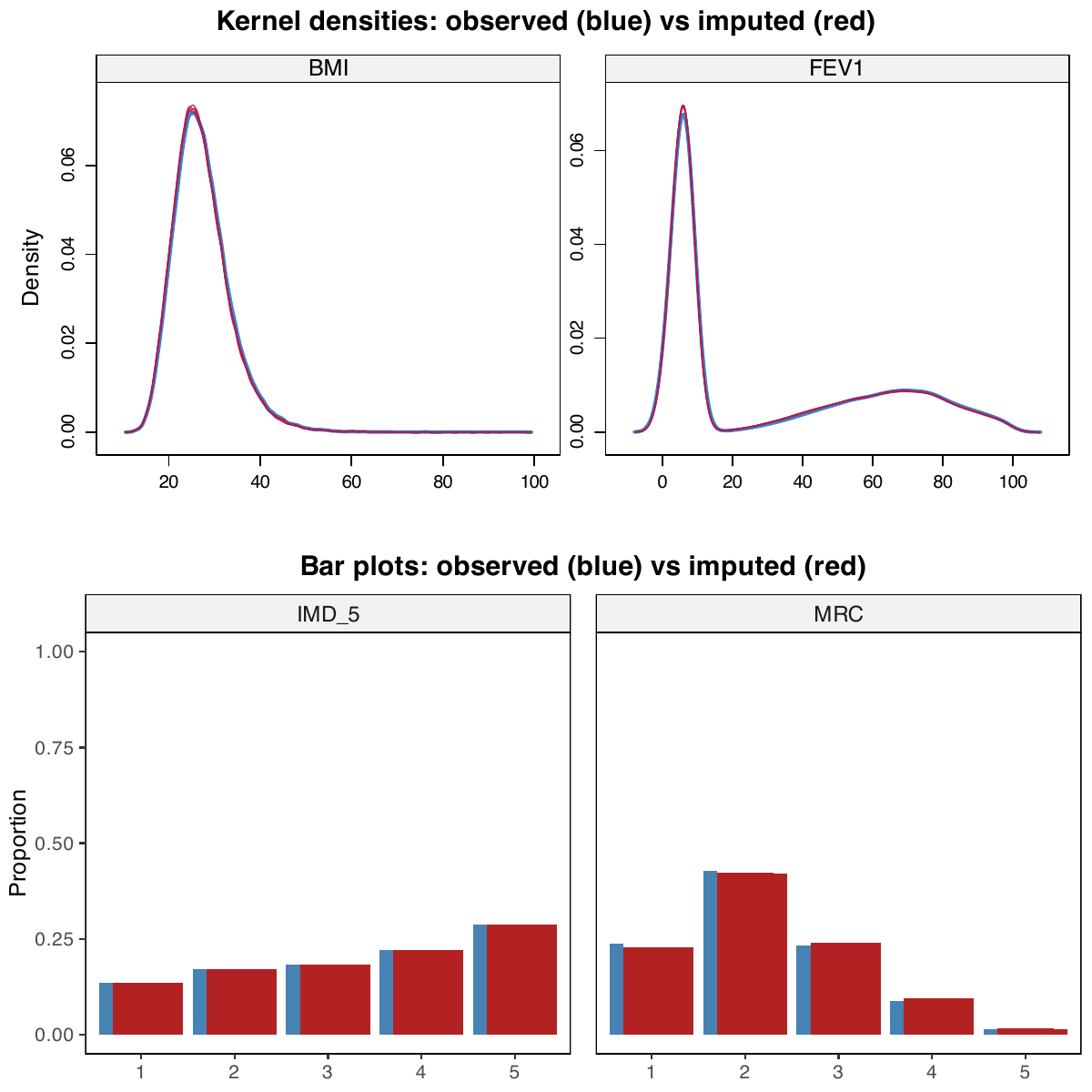}
    \caption{Kernel density estimates and bar plots of the marginal distributions of the observed data (blue) and the five densities per variable calculated from the imputed data (red) for four selected variables.}
    \label{fig:imputation_check}
\end{figure}

\textbf{Missingness of four measures.} These are electronic health records-derived observational data, where covariates were not protocolized, resulting in smaller effective samples and high missingness for the measures shown in Table~\ref{tab:missing}. $\text{IMD}_5$ and MRC are coded as categorical variables with five levels. BMI and $\text{FEV}_1$ are treated as quantitative variables, with observed ranges of (12.02, 97.90) and (0.00, 99.72), respectively.

\begin{table}[th]
\caption{Missing rate for $\text{IMD}_5$, BMI, $\text{FEV}_1$, MRC, respectively}
\label{tab:missing}
    \centering
    \begin{adjustbox}{width=0.6\linewidth}
\begin{tabular}{lllll}
\toprule
&IMD\_5 & BMI & FEV1 & MRC\\
\midrule
Missing rate &0.14\% & 58.8\%  & 55.6\% & 38.8\%\\
\bottomrule
\end{tabular}
\end{adjustbox}
\end{table}

The algorithm \fun{mice} iteratively imputes the missing values for each variable.
The imputed values randomly drawn from observed data based on predictions from fitted regressions.
The predictive mean matching method is used to impute the numeric variables BMI and $\text{FEV}_1$, while the proportional odds logistic regression method is used to impute the ordered categorical variables $\text{IMD}_5$ and MRC. These two methods correspond to `pmm' and `polr' as implemented in the \pkg{mice} package (version 3.17.0). The imputed variables were predicted from all non-missing variables, with dates transformed into numerical day counts and then rescaled into years to avoid excessively large values, and with character variables converted into factors.

We set the number of imputations to five and checked whether the imputations generated by the \fun{mice} algorithm were plausible. 
For numeric variables, we plotted the observed and imputed values to assess reasonableness. 
For categorical variables, we compared the observed and imputed distributions using bar graphs. 
Figure~\ref{fig:imputation_check} shows that the densities and bar plots of the observed and imputed values are similar. Here, we randomly selected one of the five imputed datasets to use as the working dataset. This is not the optimal approach for handling missing data and the uncertainty introduced by imputation. A better strategy would be to use all five imputations and pool the results; however, due to additional computational requirements and complications, we leave this aside for now.

\subsection*{Appendix E. Convergence diagnostics}
Because the moderate calibration curve is estimated from the MCMC samples, we assess convergence by examining whether different initialization points yield a similar estimate of the metric.
However, unlike the \pkg{dbarts} package, the \pkg{BART} package used in this work does not allow direct control over sampler configuration and does not provide a direct mechanism for users to specify custom initialization points, since these are determined internally.
In particular, for the implementation of binary probit, the model is centered around a working baseline, denoted by \fun{binaryOffset}, so the trees are initialized as deterministic null trees with terminal value zero.
Consequently, the initial random starting point arises from seeded draws of the latent variables from a truncated standard normal distribution.

\begin{figure}[h]
    \centering
    \includegraphics[width=1\linewidth]{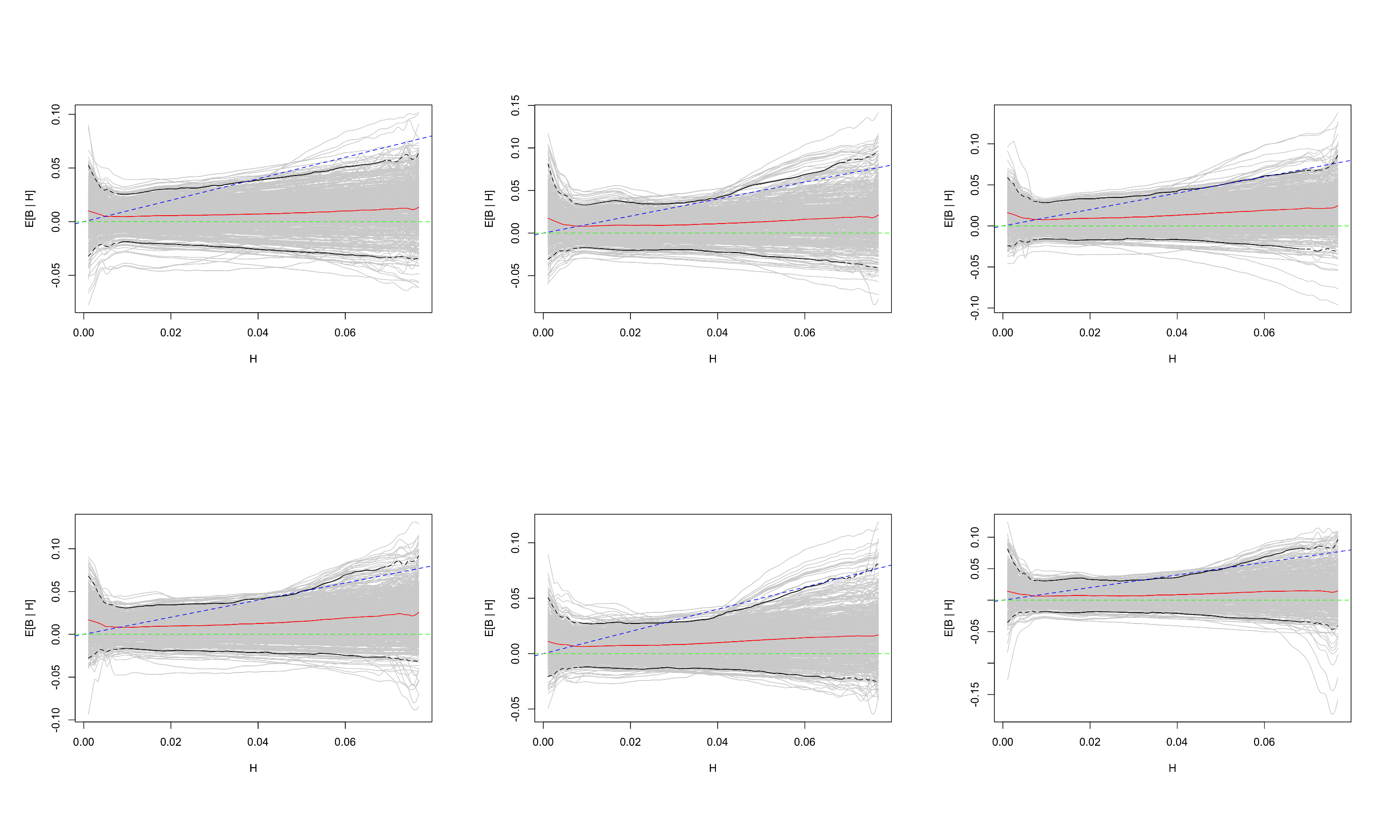}
    \caption{Moderate calibration curves with different starting points obtained by varying the random seed using the \pkg{BART} package.}
    \label{fig:seeds}
\end{figure}

Here, we varied the random seed to induce different internal initializations rather than specifying starting values directly. Figure~\ref{fig:seeds} illustrates 500 draws of the moderate calibration curve obtained using six different seeds: 11, 22, 33, 44, 55 and 66. 
The figure shows that the different chains yield similar metric estimates, which supports local stability. However, this convergence check is weaker than the evidence obtained from intentionally dispersed initializations because these induced starting points may not be widely separated.
Therefore, agreement across runs should be interpreted as supportive, rather than definitive, evidence of convergence.
For example, it cannot exclude the possibility that more distant starting states may behave differently, especially when the posterior distribution is multimodal or the sampler is sticky.

\end{document}